\documentclass[twocolumn]{aastex62}

\usepackage{hyperref}
\usepackage{amsmath}
\usepackage{subcaption}

\newcommand{\Hi}{\ion{H}{1}}

\newcommand{\Nii}{[\ion{N}{2}]}

\newcommand{\Oiii}{[\ion{O}{3}]}
\newcommand{\Sii}{[\ion{S}{2}]}
\newcommand{\Siii}{[\ion{S}{3}]}

\newcommand{\Ha}{H$\alpha$}
\newcommand{\Hb}{H$\beta$}

\newcommand{\kms}{\ifmmode\,{\rm km}\,{\rm s}^{-1}\else km$\,$s$^{-1}$\fi}
\newcommand{\um}{$\mathrm{\mu}$m}
\newcommand{\Msun}{\mathrm{M}_{\sun}}
\newcommand{\sfrunit}{\mathrm{M}_{\sun}~\mathrm{yr}^{-1}}

\defcitealias{Dopita16}{D16}

\graphicspath{{./}{}}

\shorttitle{MSA-3D survey description and first results}
\shortauthors{Bari\v{s}i\'{c} et al.}

\begin{document}

% \title{Reducing systematic uncertainties in strong-line metallicity
% measurements:  a case study in a low mass galaxy at $z\sim1$}

\title{MSA-3D: dissecting galaxies at $z\sim1$ with high spatial and spectral resolution}

\correspondingauthor{Ivana Bari\v{s}i\'{c}}
\email{ibarisic@ucdavis.edu}

\author[0000-0001-6371-6274]{Ivana Bari\v{s}i\'{c}}
\affiliation{Department of Physics and Astronomy, University of California, Davis, 1 Shields Avenue, Davis, CA 95616, USA}

\author[0000-0001-5860-3419]{Tucker Jones}
\affiliation{Department of Physics and Astronomy, University of California, Davis, 1 Shields Avenue, Davis, CA 95616, USA}

\author[0000-0001-9676-5005]{Kris Mortensen}
\affiliation{Department of Physics and Astronomy, University of California, Davis, 1 Shields Avenue, Davis, CA 95616, USA}

\author[0000-0003-2804-0648]{Themiya Nanayakkara}
\affiliation{Centre for Astrophysics and Supercomputing, Swinburne University of Technology, Hawthorn, VIC 3122, Australia}

\author[0000-0003-4520-5395]{Yuguang Chen}
\affiliation{Department of Physics and Astronomy, University of California, Davis, 1 Shields Avenue, Davis, CA 95616, USA}

\author[0000-0003-4792-9119]{Ryan Sanders}
\affiliation{Department of Physics and Astronomy, University of Kentucky, 505 Rose Street, Lexington, KY 40506, USA}
\affiliation{Department of Physics and Astronomy, University of California, Davis, 1 Shields Avenue, Davis, CA 95616, USA}

\author[0000-0003-4298-5082]{James S. Bullock}
\affiliation{Department of Physics and Astronomy, University of California Irvine, Irvine, CA 92697, USA}

\author[0000-0001-9742-3138]{Kevin Bundy}
\affiliation{Department of Astronomy \& Astrophysics, University of California, Santa Cruz, 1156 High St, CA 95064, USA}

\author[0000-0002-4900-6628]{Claude-Andr{\' e} Faucher-Gigu{\` e}re}
\affiliation{Department of Physics \& Astronomy and CIERA, Northwestern University, 1800 Sherman Ave, Evanston, IL 60201, USA}

\author[0000-0002-3254-9044]{Karl Glazebrook}
\affiliation{Centre for Astrophysics and Supercomputing, Swinburne University of Technology, Hawthorn, VIC 3122, Australia}

\author[0000-0002-6586-4446]{Alaina Henry}
\affiliation{Space Telescope Science Institute, 3700 San Martin Drive, Baltimore, MD 21218, USA}
\affiliation{Department of Physics and Astronomy, Johns Hopkins University, Baltimore, MD 21218, USA}

\author[0000-0002-5815-2387]{Mengting Ju}
\affiliation{School of Astronomy and Space Science, University of Chinese Academy of Sciences (UCAS), Beijing 100049, China}

\author[0000-0001-6919-1237]{Matthew Malkan}
\affiliation{Department of Physics and Astronomy, University of California Los Angeles, 430 Portola Plaza, Los Angeles, CA 90095, USA}

\author[0000-0002-8512-1404]{Takahiro Morishita}
\affiliation{IPAC, California Institute of Technology, MC 314-6, 1200 E. California Boulevard, Pasadena, CA 91125, USA}

\author[0000-0002-1527-0762]{Danail Obreschkow}
\affiliation{International Centre for Radio Astronomy Research (ICRAR), M468, University of Western Australia, Perth, WA 6009, Australia}
\affiliation{Australian Research Council, ARC Centre of Excellence for All Sky Astrophysics in 3 Dimensions (ASTRO 3D), Australia}

\author[0000-0002-4430-8846]{Namrata Roy}
\affiliation{Center for Astrophysical Sciences, Department of Physics and Astronomy,
Johns Hopkins University, Baltimore, MD 21218, USA}

\author[0000-0001-6703-4676]{Juan M. Espejo Salcedo}
\affiliation{Max-Planck-Institut für extraterrestische Physik (MPE), Giessenbachstr., 85748 Garching, Germany}

\author[0000-0003-3509-4855]{Alice E. Shapley}
\affiliation{Department of Physics and Astronomy, University of California Los Angeles, 430 Portola Plaza, Los Angeles, CA 90095, USA}

\author[0000-0002-8460-0390]{Tommaso Treu}
\affiliation{Department of Physics and Astronomy, University of California Los Angeles, 430 Portola Plaza, Los Angeles, CA 90095, USA}

\author[0000-0002-9373-3865]{Xin Wang}
\affiliation{School of Astronomy and Space Science, University of Chinese Academy of Sciences (UCAS), Beijing 100049, China}
\affiliation{National Astronomical Observatories, Chinese Academy of Sciences, Beijing 100101, China}
\affiliation{Institute for Frontiers in Astronomy and Astrophysics, Beijing Normal University, Beijing 102206, China}

\author[0000-0003-1809-6920]{Kyle B. Westfall}
\affiliation{University of California Observatories, University of California, Santa Cruz, 1156 High St., Santa Cruz, CA 95064, USA}

%\author{Alice Shapley}



\begin{abstract}

Integral field spectroscopy (IFS) is a powerful tool for understanding the formation of galaxies across cosmic history. We present the observing strategy and first results of MSA-3D, a novel JWST program using multi-object spectroscopy in a slit-stepping strategy to produce IFS data cubes. The program observed 43 normal star-forming galaxies at redshifts $0.5 \lesssim z \lesssim 1.5$, corresponding to the epoch when spiral thin-disk galaxies of the modern Hubble sequence are thought to emerge, obtaining kpc-scale maps of rest-frame optical nebular emission lines with spectral resolution $R\simeq2700$. Here we describe the multiplexed slit-stepping method which is $>15$ times more efficient than the NIRSpec IFS mode for our program. As an example of the data quality, we present a case study of an individual galaxy at $z=1.104$ (stellar mass $M_* = 10^{10.3}~\Msun$, star formation rate~$=3~\Msun$\,yr$^{-1}$) with prominent face-on spiral structure. We show that the galaxy exhibits a rotationally supported disk with moderate velocity dispersion ($\sigma = 36^{+5}_{-4}$~\kms), a negative radial metallicity gradient ($-0.020\pm0.002$~dex\,kpc$^{-1}$), a dust attenuation gradient, and an exponential star formation rate density profile which closely matches the stellar continuum. These properties are characteristic of local spirals, indicating that mature galaxies are in place at $z\sim1$. We also describe the customized data reduction and original cube-building software pipelines which we have developed to exploit the powerful slit-stepping technique. Our results demonstrate the ability of JWST slit-stepping to study galaxy populations at intermediate to high redshifts, with data quality similar to current surveys of the $z\sim0.1$ universe.

%Reliable kinematic data during galaxies’ formative epochs is crucial for understanding the transition from irregular clump structures to thin disks. Additionally, spatially resolved attenuation and star formation profiles are vital probes of galaxy growth and structural evolution. %Integral field spectroscopy has proven instrumental in characterizing aforementioned galaxy properties.
%GO-2136 NIRSpec/MSA slit-stepping program is first to provide high resolution kpc-scale pseudo-IFS observations at the epoch of modern thin disks emergence (0.5$<$z$<$1.5). We sample rest-optical emission lines for 43 main-sequence galaxies. Here, we detail a case study for an individual galaxy selected based on its morphology — a distinct spiral disk structure. We find: a) Kinematic data are well-fit by a disk model, indicating a smooth rotational curve with evident rotational support. b) Characterization of gas-phase metallicity and dust attenuation gradients utilizing spatially resolved maps of emission line ratios reveals clear negative radial trends.
%We outline the MSA slit-stepping technique: in only $\sim$2hrs/source it achieves the sensitivity an equivalent IFU survey would demand 8hrs/source for, making MSA slit-stepping over 15 times more efficient. Customized data reduction pipelines and original software for cube building were developed to handle the program's complexity, as slit-stepping lacks standard pipeline support, necessitating substantial development efforts detailed in this paper.

\end{abstract}


\keywords{Galaxy formation (595), Galaxy evolution (594), Disk galaxies (391), High-redshift galaxies (734), Astronomical techniques (1684)}


\section{Introduction} \label{sec:intro}

Galaxies in the nearby universe can be classified according to the ``Hubble sequence'' of star-forming disks with spiral arms, and passive spheroids \citep[e.g.,][]{hubble1926,kennicutt1998}. A similar sequence persists at cosmological redshifts of at least $z\gtrsim3$ \citep[e.g.,][]{wuyts2011,mortlock2013,zhang2019,straatman2016}. Integral field spectroscopy (IFS) has proven critical in understanding the structure of disk galaxies in the early universe, in particular by observing strong emission lines such as \Ha\ and CO transitions \citep[e.g., see reviews by][]{glazebrook2013,forster2020}. 
IFS surveys have revealed that disk galaxies at $z\gtrsim2$ are characterized by higher velocity dispersion $\sigma$ (and lower ratio of rotation to dispersion support, $V/\sigma$; e.g., \citealt{ubler2019,leethochawalit2016,simons2016,epinat2012,jones2021_alpine}), scale heights characteristic of thick disks \citep[e.g.,][]{elmegreen2006,hamilton2023}, shallower gas-phase metallicity gradients \citep[indicating radial mixing; e.g.,][]{wang2019,wang2020}, and clumpier star formation morphology \citep{forster2011,guo2015,livermore2015}, typically lacking the spiral structure seen at low redshifts. 
%CAFG editing/expanding on theory comments
%Cosmological simulations indicate that these differences arise from higher accretion rates, with higher gas fractions and disk instability, leading to bursty star formation and increased turbulence at higher redshifts \citep[e.g.,][]{dekel2009,brennan2017,pillepich2019,forbes2023,gurvich2023}. The emergence of modern thin disk galaxies with spiral structure is thus thought to be linked to a decline in merger rates and accretion from the intergalactic and circumgalactic medium, coupled with concentration of the gravitational potential \citep[e.g.,][]{hopkins2023}. 

Different factors contributing to the higher gas velocity dispersions at higher redshifts have been proposed theoretically. 
One effect is that higher gas accretion rates result in higher gas fractions in galaxies, which in equilibrium disk models correlate with higher velocity dispersions \citep[e.g.,][]{dekel2009,fg2013,brennan2017,pillepich2019,forbes2023}. 
Recently, high-resolution cosmological simulations have also found that the thermodynamic state of the circumgalactic medium (CGM) correlates tightly with the emergence of thin disks that can support spiral structure. 
In particular, \cite{hafen2022} showed that a virialized inner CGM allows the gas to accrete with coherent angular momentum onto central galaxies via hot-mode accretion, which promotes the formation of thin disks. 
\cite{stern2021} also argued that the high and near-uniform thermal pressure in a virialized inner CGM helps confine star formation-driven outflows. 
In a related study, \cite{hopkins2023} emphasized the role of the concentration and depth of the gravitational potential. 
The former is a determinant of whether there is a well-defined dynamical center around which rotation can stabilize, while the latter (which correlates with CGM virialization) contributes to the confinement of outflows via gravity \citep[see][for a discussion of outflow confinement by hot gas pressure vs. gravity]{byrne2023}. 
In this picture in which the emergence of thin disks is connected to the virialization of the inner CGM and the depth of the gravitational potential, mass rather than redshift is a primary predictor of thin-disk formation. In practice, this leads to an increased thin-disk fraction going to lower redshift because the abundance of sufficiently massive halos ($M_{\rm halo} \gtrsim {\rm a~few } \times 10^{11}$ M$_{\odot}$) increases with decreasing redshift.

Observationally, the epoch around $z\simeq1$ is thought to be a time of ``disk settling'' during which thin spiral disks become prominent among star forming galaxies \citep[e.g.][]{kassin2012, miller2011, miller2012, simons2017}.

 % The epoch \textcolor{magenta}{around} $z\simeq1$, when the universe was approximately half of its present age, is thought to be a time of ``disk settling'', \textcolor{magenta}{during which galaxies form thin spiral disks} %when galaxies develop thin spiral disks 
 % \citep[e.g.][]{kassin2012, miller2011, miller2012, simons2017}. 
Surveys of galaxy kinematics indicate a rise in the degree of rotational support revealed by increasing $V/\sigma$ toward lower redshifts \citep[as well as higher $V/\sigma$ at higher masses; e.g.,][]{wisnioski2019,simons2016,tiley2021}. For example, \cite{simons2017} show that the fraction of massive galaxies (stellar mass $M_* > 10^{9.5}~\Msun$) with $V/\sigma > 3$ increases from $\sim$20\% at $z=2$ to approximately 50\% at $z=1$. Morphological studies similarly show an increasing fraction of disk and spiral galaxies at lower redshifts. While clear examples of spiral galaxies have been found as early as $z\gtrsim3$ \citep[e.g.,][]{wu2023}, they are rare at such redshifts. Recent analysis of James Webb Space Telescope (JWST) imaging by \cite{kuhn2024} found that the fraction of massive galaxies ($M_* > 10^{10}~\Msun$) with visually identifiable spiral structure increases from $\sim$15\% at $z=2$ to 43\% at $z=1$, although the true spiral fraction is likely higher given the challenge of detecting spiral structure at high redshifts. Previous investigations of Hubble Space Telescope images found broadly similar results \citep[e.g.,][]{margalef-bentabol2022}. 
Overall, morphological and kinematic properties both indicate that well-ordered disks are more common at lower redshift, with the population of moderately massive star-forming galaxies becoming dominated by spiral disk morphologies at around $z\simeq1$. 

High-quality IFS observations of strong nebular emission lines (such as \Ha, \Nii, and \Oiii) can provide a comprehensive view of individual high-redshift galaxies and their evolutionary pathways. For example, kpc-scale kinematic measurements can reveal the mass density profile and dark matter fraction \citep[e.g.,][]{genzel2020}. 
In addition to kinematics and morphology, strong emission line ratios provide information about ionization sources \citep[e.g.,][]{wright2010,belli2017} and gas-phase metallicity \citep[e.g.,][]{maiolino2019}. Radial metallicity gradients have emerged as a valuable probe of the ``baryon cycle'' of gas accretion and feedback-driven outflows which regulate galaxy formation \citep[][]{tumlinson2017,fg2023}, as established by various theoretical investigations \citep{gibson2013,ma2017,hemler2021}. 
Metallicity and kinematic maps can also identify signatures of metal-poor gas accretion \citep[e.g.,][]{sanchez_almeida2015,ju2022}. 
Observations of high-redshift galaxies have found a large scatter in gradient slopes, which are shallower on average than in nearby spiral galaxies and show substantial azimuthal variation \citep[e.g.,][]{wang2017,wang2020,wang2022,curti2020}. 
As with kinematics, spatial resolution of $\lesssim1$~kpc is essential for accurate results \citep{yuan2013,jones2013}. 
IFS emission line maps can thus not only identify disks via their kinematic signatures, but also probe the properties such as accretion and gravitational potential, which allow the formation of thin spiral disks.

A key challenge for high redshift galaxies is to obtain IFS with sufficient angular resolution, as well as spectral resolution and signal-to-noise ratio. 
For the rest-frame optical emission lines, the vast majority of current samples are observed with seeing-limited data of $\sim$0\farcs5--1\farcs0 resolution. This corresponds to $\sim$4--8 kpc at $z\gtrsim1$ \citep[e.g.,][]{forster2009,wisnioski2015,wisnioski2019,stott2016} which is comparable to massive galaxy sizes at these redshifts. 
Such data cannot clearly distinguish disks from merging systems at moderate to high redshifts \citep[e.g.,][]{leethochawalit2016,simons2019}. 
Instead, the prevalence and kpc-scale structure of rotationally supported disks at redshifts $z\gtrsim2$ has been established in smaller samples with adaptive optics (AO) assisted observations \citep[e.g.,][]{genzel2006,genzel2008,forster2018,espejo2022}, including AO surveys of gravitationally lensed galaxies which can probe $\sim$100 parsec scales \citep{stark2008,jones2010,leethochawalit2016,hirtenstein2019}. 
However, current AO systems are most effective only at wavelengths $\gtrsim1.5$~\um. Consequently, most AO IFS samples are limited to $z\gtrsim1.5$ where \Ha\ and other strong lines can be observed with high Strehl ratios. 
As a result, the epoch of spiral disk emergence at $z\simeq0.5$--1.5 has not been explored with suitable resolution. 
Additionally, ground-based observations suffer from a time-variable atmosphere and point-spread function (PSF), which is especially complex for AO data. This leads to challenges in recovering properties such as galaxy velocity dispersions and $V/\sigma$ \citep[e.g.,][]{davies2011,burkert2016}.
We note that space-based grism observations are able to obtain kpc-scale spatial resolution but are limited by poor spectral resolution \citep[$R \lesssim 300$; e.g.,][]{simons2021,wang2020,jones2015}, while ground-based IFS surveys are limited to coarse spatial resolution at the key wavelengths. 

JWST's unique capabilities now enable excellent spatial \textit{and} spectral resolution, combined with good sensitivity at the relevant wavelengths for mapping rest-frame optical emission lines at $z\gtrsim0.5$ with the NIRSpec instrument \citep{jakobsen2022,boker2023}. 
Importantly, JWST delivers reliable results via a stable PSF with high Strehl, in contrast to ground-based AO observations.
However, NIRSpec's integral field unit (IFU) is effectively limited to single-object observations given the 3\arcsec\ field of view. Building sufficient samples to characterize the galaxy population at high redshifts is thus a time-intensive prospect. To efficiently obtain high-quality data for large samples, we have instead turned to NIRSpec's Micro-Shutter Assembly (MSA) which provides multiplexed observations with 0\farcs2 wide slitlets. By utilizing a ``slit-stepping'' observing sequence \citep[e.g.,][]{ho2017,genzel2013}, we are able to obtain IFS data for $>$40 galaxies simultaneously in $<10$\% of the time that would be required for an equivalent IFU survey. This paper describes our first application and example results of this novel multiplexed slit-stepping method via the program JWST-GO-2136, in which we observed 43 galaxies at $0.5 < z < 1.7$, corresponding to the disk settling epoch when spiral structure becomes prominent. As we are obtaining 3-D data cubes using NIRSpec's MSA strategy, we dub our program MSA-3D\footnote{With a nod to previous surveys 3D-HST \citep{brammer2012} and KMOS$^{\mathrm{3D}}$ \citep{wisnioski2015}.}.

In this paper, we demonstrate that our technique provides excellent data quality for mapping $z\sim1$ galaxies, which our team is analyzing with a series of planned papers on their kinematics, metallicity gradients, excitation sources, star formation morphology, and more. 
Notably, our sample size obtained with a $\simeq$30 hour MSA slit-stepping program is comparable to the 47 galaxies targeted by the 273-hour NIRSpec guaranteed time GA-NIFS program using the IFU \citep{perna2023}. 
Given the large efficiency gain over JWST's IFU mode, our method represents the most pragmatic opportunity to build large samples (tens to hundreds) of kpc-resolution IFS data of distant galaxies to study their evolution across cosmic time. Accordingly, one of our objectives in this paper is to simply demonstrate the effectiveness of this observing strategy. We urge its adoption for large multiplexed galaxy surveys with JWST. 

This paper describes the observational strategy and data reduction procedure, serving as a reference for the MSA-3D program and subsequent papers. We also describe a case study of an individual galaxy illustrating the variety of analyses enabled by MSA-3D. 
Various analyses of the broader sample are in preparation or planned. These include a companion paper by Ju et al. (in prep.) presenting the gas-phase metallicity gradients, a study of the gas ionization mechanisms and active galactic nuclei (Roy et al. in prep.), and other topics such as kinematics, angular momentum, dark matter content, and star formation rate profiles.
The structure of the present paper is as follows. In Section~\ref{sec:Data_Acquisition} we describe the JWST observations, including the sample selection, motivation and implementation of the slit-stepping methodology, and data processing. Section~\ref{sec:results} presents a case study of a $z=1.104$ spiral galaxy based on emission line measurements from our data. We describe its kinematic structure, gas-phase metallicity and radial gradient, dust attenuation and radial gradient, and overall disk structure as traced by the surface brightness profile. In Section~\ref{sec:discussion} we discuss our results, and future analyses enabled by the MSA-3D program, in the context of broad efforts to understand the growth and evolution of galaxies across cosmic history. 
Throughout this work we adopt a flat $\Lambda$CDM cosmology with H$_0 = 69.6$~km\,s$^{-1}$\,Mpc$^{-1}$ and $\Omega_M = 0.286$ \citep{bennett2014}.

\begin{figure*}[ht]
    \centering
    \includegraphics[width=\columnwidth]{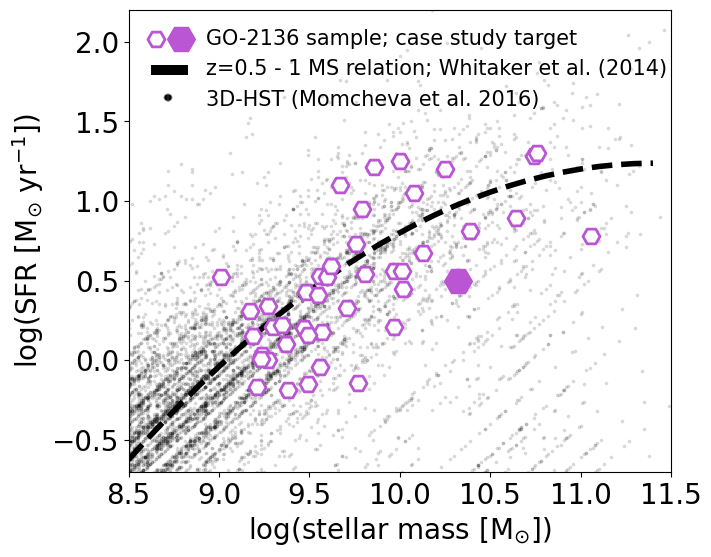}
    \includegraphics[width=0.91\columnwidth]{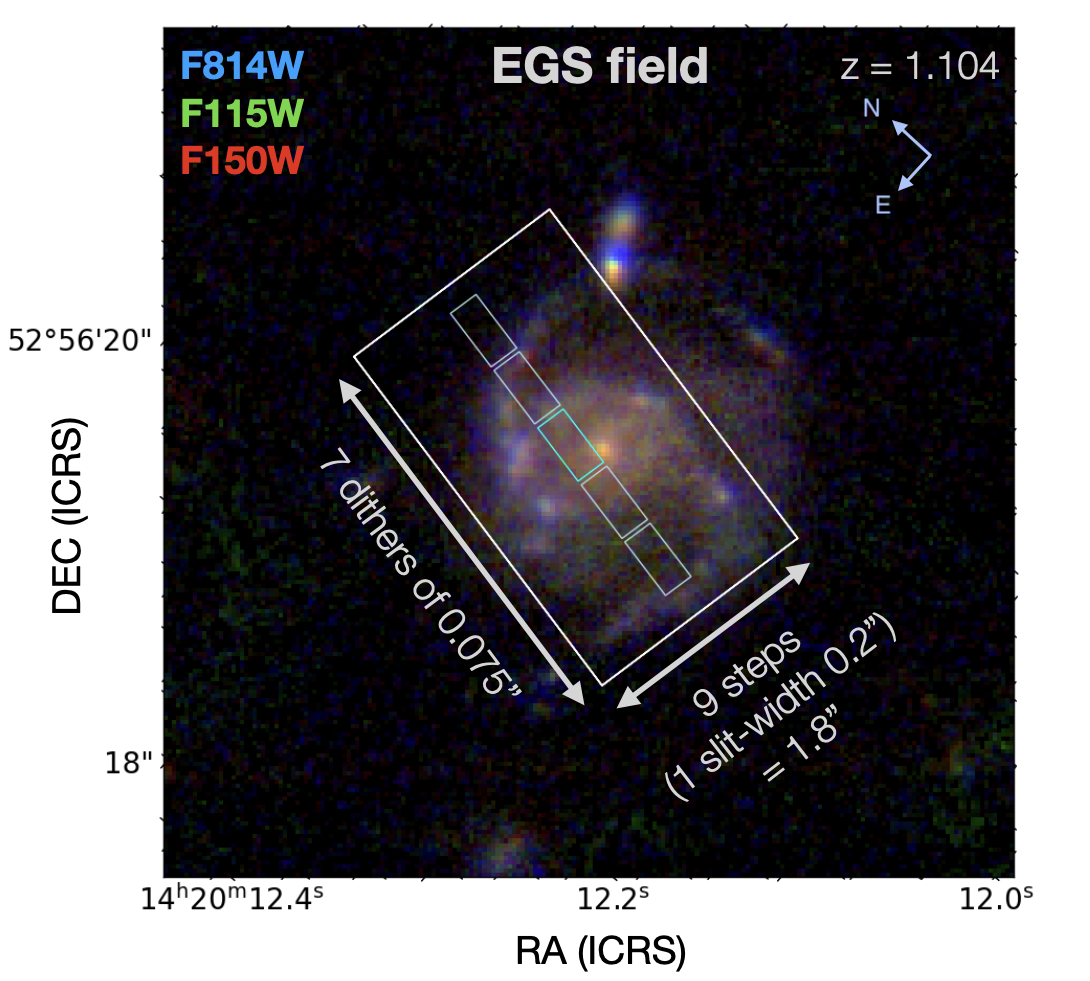}
    \caption{\textit{Left:} 
    The MSA-3D target sample from program JWST-GO-2136 (purple hexagons) is selected to represent typical star-forming galaxies within the redshift range $z=0.5$--1.7. We show their star formation rates and stellar masses, along with the full 3D-HST sample within the same redshift range. 
    The black dashed line depicts the star forming main sequence relation from \cite{whitaker2014}. Our targets are representative of the star-forming population at stellar masses $\gtrsim 10^9~\Msun$.
    \textit{Right:} Color image of the $z=1.1$ galaxy analyzed in this paper (ID~8512; blue: HST/ACS F814W, green: JWST/NIRCam F115W, red: JWST/NIRCam F150W). The galaxy shows a clear face-on spiral morphology. 
    % with a redder central bulge. 
    This target was observed with five NIRSpec/MSA slitlets, shown as small grey rectangles. Arrows show the slit-stepping approach with 9 steps of 0\farcs2 in the dispersion direction, and 7 dithers of 0\farcs075 in the cross dispersion direction. Cross-dispersion steps are used to remove the bar shadows (0\farcs075 width) which can be seen as the gaps between slitlets. The larger rectangle shows the full 1\farcs8$\times$3\farcs0 sky area of the resulting IFS data cube, of which each pixel in the central 1\farcs8$\times$2\farcs1 has uniform coverage of 117 minutes from six exposures on-source (and a seventh exposure affected by bar shadows). 
    }
    \label{fig:color_image}
\end{figure*}

\begin{table*}[t]
  \footnotesize
  \caption{Properties of the target sample. The ID numbers, stellar masses, and star formation rates are taken from the 3D-HST catalogs \citep{brammer2012,skelton2014}. For each object we list the strong nebular emission lines covered by the NIRSpec/MSA spectra. \Nii, \Sii, \Oiii, and \Siii\ refer to both lines in the doublet unless specified otherwise.} 
  \tabcolsep=0.08cm
  \label{tab:sample}
  \centering
\begin{tabular}{cccccccc}
\hline\hline
ID	     &	 RA	  &	 Dec  &	$z$  &	stellar mass & SFR & emission lines  \\
    &	 	&	 	    	&	         	& $\rm log(M_{*}/M_{\odot})$   & $\rm M_{\odot} yr^{-1}$   &  \\
\hline
2111	&	215.0627824	&	52.9070766	&	0.58 	&	9.97 	&	1.6 	&	\Ha, \Nii, \Sii, \Siii	\\
2145	&	215.0694675	&	52.9108516	&	1.17 	&	9.19 	&	1.4 	&	\Ha, \Nii, \Sii, \Hb, \Oiii 	\\
2465	&	215.0704381	&	52.9137241	&	1.25 	&	9.30 	&	1.6 	&	\Ha, \Nii, \Sii, \Hb, \Oiii		\\
2824	&	215.0685025	&	52.914333	&	0.98 	&	9.49 	&	0.7 	&	\Ha, \Nii, \Sii, \Oiii, \Siii~$\lambda$9069 	\\
3399	&	215.042511	&	52.8996031	&	1.34 	&	9.81 	&	3.5 	&	\Ha, \Nii, \Sii, \Hb, \Oiii		\\
4391	&	215.0676175	&	52.9232437	&	1.08 	&	9.48 	&	2.7 	&	\Ha, \Nii, \Hb, \Oiii 	\\
6199	&	215.0450158	&	52.9194652	&	1.59 	&	10.00 	&  17.8 	&	\Ha, \Nii, \Sii, \Hb 	\\
6430	&	215.0131444	&	52.8980378	&	1.17 	&	9.79 	&	8.9 	&	\Ha, \Nii, \Sii, \Hb, \Oiii~$\lambda$4959 	\\
6848	&	215.0355588	&	52.9166925	&	1.57 	&	10.64 	&	7.8 	&	\Ha, \Nii, \Sii, \Oiii		\\
7314	&	214.9989097	&	52.8925151	&	1.28 	&	9.55 	&	2.6 	&	\Ha, \Nii, \Hb, \Oiii 	\\
7561	&	215.0609094	&	52.9383828	&	1.03 	&	9.21 	&	0.7 	&	\Ha, \Nii, \Sii, \Hb, \Oiii, \Siii~$\lambda$9069		\\
8365	&	215.0599904	&	52.9422373	&	1.68 	&	9.56 	&	3.4 	&	\Ha, \Nii, \Sii, \Hb, \Oiii 	\\
8512	&	215.0497806	&	52.9380795	&	1.10 	&	10.32 	&	3.2 	&	\Ha, \Nii, \Sii, \Hb, \Oiii 	\\
8576	&	215.0595679	&	52.9434335	&	1.57 	&	9.60 	&   3.3 	&	\Ha, \Nii, \Sii, \Hb, \Oiii 	\\
8942	&	215.0094032	&	52.9100655	&	1.18 	&	9.86 	&  16.2 	&	\Ha, \Nii, \Sii, \Hb, \Oiii~$\lambda$4959  	\\
9337	&	214.9957065	&	52.9019407	&	1.17 	&	9.27 	&	2.2 	&	\Ha, \Nii, \Sii, \Hb, \Oiii		\\
9424	&	214.9926533	&	52.900911	&	0.98 	&	9.76 	&	5.4 	&	\Ha, \Nii, \Sii 	\\
9482	&	215.0530158	&	52.9442406	&	1.21 	&	9.77 	&	0.7 	&	\Ha, \Nii \Sii, \Hb, \Oiii		\\
9527	&	215.0085005	&	52.9123786	&	1.42 	&	10.02 	&	2.8 	&	\Ha, \Nii, \Hb, \Oiii		\\
9636	&	215.0365985	&	52.9328783	&	0.74 	&	9.35 	&	1.7 	&	\Ha, \Nii, \Sii, \Siii 	\\
9812	&	215.0402843	&	52.9376004	&	0.74 	&	9.97 	&	3.6 	&	\Ha, \Nii, \Sii, \Siii 	\\
9960	&	215.031894	&	52.9331513	&	1.51 	&	11.06 	&	6.0 	&	\Ha, \Nii, \Sii 	\\
10107	&	214.9817759	&	52.8975743	&	1.01 	&	10.01 	&	3.6 	&	\Ha, \Nii, \Sii, \Oiii 	\\
10502	&	214.9857711	&	52.9033048	&	1.23 	&	10.13 	&	4.7 	&	\Ha, \Nii, \Hb, \Oiii 	\\
10752	&	215.0403761	&	52.9413773	&	1.73 	&	9.62 	&	3.9 	&	\Hb, \Oiii		\\
10863	&	215.0551264	&	52.9529908	&	1.03 	&	9.24 	&	1.1 	&	\Ha, \Nii, \Hb, \Oiii, \Siii~$\lambda$9069 	\\
10910	&	215.0561932	&	52.9553701	&	0.74 	&	9.57 	&	1.5 	&	\Ha, \Nii, \Sii, \Siii 	\\
11225	&	215.0415788	&	52.9454655	&	1.05 	&	9.67 	&  12.6 	&	\Ha, \Nii, \Sii, \Hb, \Oiii 	\\
11539	&	214.9819651	&	52.9051336	&	1.61 	&	10.39 	&	6.5 	&	\Hb, \Oiii		\\
11702	&	214.9794086	&	52.9031601	&	1.23 	&	9.37 	&	1.3 	&	\Hb, \Oiii		\\
11843	&	215.0390468	&	52.9471037	&	1.46 	&	10.74 	&	19.0 	&	\Ha, \Nii, \Sii, \Hb, \Oiii		\\
11944	&	215.0369901	&	52.9453937	&	1.04 	&	9.27 	&	1.0 	&	\Ha, \Nii, \Sii, \Hb, \Oiii 	\\
12015	&	215.0323151	&	52.9431798	&	1.24 	&	10.08 	&	11.2 	&	\Ha, \Nii, \Sii, \Hb, \Oiii 	\\
12071	&	215.0219665	&	52.9360567	&	1.28 	&	9.47 	&	1.6 	&	\Ha, \Nii, \Sii, \Hb, \Oiii		\\
12239	&	215.0495343	&	52.9560252	&	0.89 	&	9.49 	&	1.5 	&	\Ha, \Nii, \Sii, \Siii 	\\
12253	&	215.0442096	&	52.9520815	&	1.03 	&	9.17 	&	2.0 	&	\Sii, \Hb, \Oiii\\
12773	&	215.029765	&	52.9451588	&	0.95 	&	9.23 	&	1.0 	&	\Ha, \Nii, \Sii 	\\
13182	&	214.99983	&	52.9268186	&	1.54 	&	10.25 	&  15.8 	&	\Hb, \Oiii		\\
13416	&	215.0252815	&	52.9456859	&	1.54 	&	10.76 	&	20.0 	&	\Ha, \Nii, \Sii, \Oiii 	\\
18188	&	214.9839579	&	52.9413562	&	0.82 	&	9.56 	&	0.9 	&	\Ha, \Nii, \Sii 	\\
18586	&	214.9712282	&	52.9337862	&	0.76 	&	9.01 	&	3.3 	&	\Ha, \Nii, \Sii 	\\
19382	&	214.9765628	&	52.9414977	&	1.03 	&	9.38 	&	0.6 	&	\Ha, \Nii, \Sii, \Oiii	\\
29470	&	214.9689505	&	52.9453733	&	1.04 	&	9.71 	&	2.1 	&	\Ha, \Nii, \Sii, \Hb, \Oiii 	\\
\hline
\end{tabular}\\

% Notes:
% $^a$ From the UVCANDELS catalog, using Dense Basis SED modeling based on HST observations ranging from F275W to F160W filters.\\
% $^b$ Gaussian-fitted emission lines: \Nii\ is \Nii\lambda6548,6584, \Sii\ is \Sii\lambda6717,6731, and \Oiii\ is \Oiii\lambda4959,5007.
\end{table*}

\section{Observations and Data Processing}
\label{sec:Data_Acquisition}

This paper is based on spectroscopic data from the JWST Cycle 1 program GO-2136 (PI: Jones). This program obtained NIRSpec Micro-Shutter Assembly (MSA) observations of 43 galaxies at $z\sim1$, using a slit-stepping strategy %mode 
to provide spatially resolved kpc-scale pseudo-integral field spectroscopy. Data were taken on 2023 March 29-30. We used the G140H/F100LP grating/filter combination, covering wavelengths 0.97--1.82~\um\ with full-width at half maximum (FWHM) spectral resolution $\Delta \lambda \simeq 5.2$~\AA\ corresponding to $R\simeq2700$ resolving power. 
As described below, dithered exposures cover a contiguous 1\farcs8$\times$(2\farcs0--3\farcs0) field of view for each target galaxy, depending on the number of slitlets used. The observations are optimized to map strong rest-frame optical emission lines in galaxies spanning the epoch in which modern thin disks emerge \citep[e.g.][]{kassin2012, miller2012, simons2017, wisnioski2019}. 
In this section, we describe the observing strategy, the merits of the MSA slit-stepping approach relative to the IFU mode, and the data reduction steps performed to produce data cubes.

\subsection{Sample selection}
\label{sec:sample}

Our target field is the Extended Groth Strip (EGS), a well-studied extragalactic field with extensive photometric and spectroscopic survey data including from the CANDELS \citep{koekemoer2011}, 3D-HST \citep{momcheva2016}, CEERS \citep{finkelstein2023}, DEEP2 and DEEP3 \citep{newman2013}, and MOSDEF \citep{kriek2015} surveys. This field was selected in part to overlap with the CEERS JWST Early Release Science imaging area. 
Our program science goals require mapping multiple nebular emission lines to study resolved kinematics, star formation rates, metallicity, and ionization mechanisms. Our primary sample selection is to target \Ha, \Nii, \Sii, \Oiii, and \Hb. These lines are accessible from $z=1.0$--1.7 using the G140H/F100LP grating/filter. We adopt a secondary target redshift range $z=0.5$--1.0 which includes coverage of \Ha, \Nii, \Sii, \Siii, and \Hi\ Paschen lines, enabling similar measurements. 
For target selection, we required spectroscopic redshift confirmation and multi-wavelength HST photometry from the CANDELS survey \citep[][]{koekemoer2011}, which, along with other imaging data, ensures reliable spectral energy distribution fitting based measurements of stellar mass and star formation rate (SFR). 
For selection purposes, we adopted stellar population parameter values from the 3D-HST survey catalogue \citep[][]{skelton2014,momcheva2016}.
In total, we identified a parent catalog of 1336 galaxies in the target redshift range $0.5 < z < 1.7$, and further restricted the MSA target selection to the 72\% of these galaxies with secure spectroscopic redshifts.
The sample is shown in the left panel of Figure~\ref{fig:color_image}.

In building our spectroscopic sample, we applied selection criteria of stellar mass $M_{*} > {10^9}~\Msun$ and SFR~$>0.6~\sfrunit$. This ensures the observed targets probe the star forming main sequence at $z\sim1$ \citep[e.g.,][]{whitaker2012,speagle2014} and have suitably bright and spatially extended nebular emission. However, we did not impose any selection criteria based on galaxy size or surface brightness to avoid biases in the observed sample. 
In designing the MSA mask, we also targeted regions with JWST/NIRcam coverage from the CEERS program. Ultimately, the MSA mask includes 43 target galaxies, of which 19 are within the current NIRCam footprint. 
Table~\ref{tab:sample} lists properties and emission line coverage of these 43 galaxies, of which 32 are at $z>1$ and 11 are at $0.5<z<1$. Coverage of the \Ha\ line is present in 38 of the 43 galaxies (88\%), whereas in the remainder of the sample it falls in the chip gap or off the detectors.
%\textcolor{orange}{Properties of these 43 galaxies and their emission line coverage are listed in Table~\ref{tab:sample}.}
%\textcolor{red}{[Check: any additional NIRCam coverage since this program was executed?]} 
%32 galaxies are at $z>1$ and 11 are at $0.5<z<1$. 
%38 of the 43 galaxies (88\%) include coverage of the \Ha\ line whereas in the remainder it falls in the chip gap or off the detectors. 
In this paper we present a case study analysis of galaxy ID~8512 at $z=1.104$ with $M_{*} = 10^{10.32}~\Msun$ and SFR~$=3.2~\sfrunit$. We selected this target for initial analysis based on its clear spiral disk morphology. %morphology, which shows a clear spiral disk structure (Figure~\ref{fig:color_image}). 
It is otherwise representative of the more massive galaxies in the observed sample. 

%\textcolor{red}{[Should we include a table of targets (ID, RA/Dec, z, lines covered)? And a figure of SFR vs stellar mass (demonstrating these are "typical SF galaxies" with decent range of mass)?]}

\subsection{Slit-stepping observing strategy}
\label{sec:observing_strategy}

\begin{figure*}
    \centering
    \includegraphics[width=\textwidth]{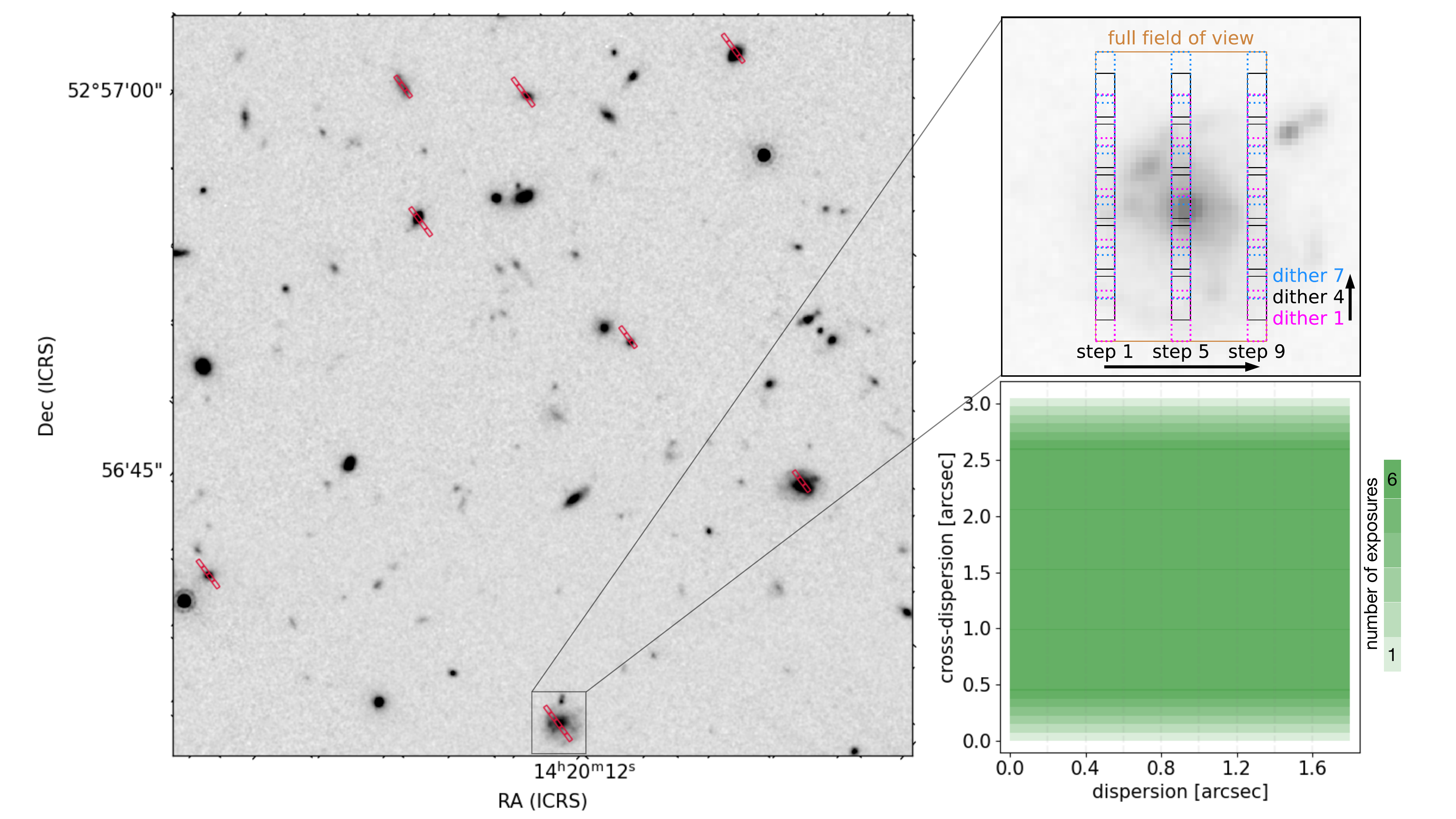}
    \caption{
    Illustration of IFS multiplexing achieved with MSA-3D's slit-stepping strategy. 
    \textit{Left:} A portion of NIRSpec/MSA's field of view. Red slitlets show the central MSA pointing for $z\sim1$ galaxies targeted by this program, overlaid on a near-infrared HST F160W image. Each galaxy is targeted with between 3--5 slitlets. 
    \textit{Top right:} Fuchsia, black, and blue boxes show slitlet positions for a subset of 9 out of the 63 pointings overlaid on a JWST/NIRCam F444W image of an example target galaxy. These represent the first, middle, and final positions in both the dispersion direction (``steps'') and cross-dispersion direction (``dithers''). Collectively the 63 pointings provide contiguous coverage over the field of view. 
    \textit{Bottom right:} Green shading shows the effective number of exposures at each position within the field, ranging from one to six. Gray dashed vertical lines denote the 0\farcs2 slit width. Our strategy achieves nearly uniform six-exposure depth over the central region, effectively eliminating the MSA bar shadows.
    This observing strategy results in IFS data for all objects on the MSA mask.
    }
    \label{fig:multiplex}
\end{figure*}

A key unique aspect of these JWST Cycle 1 observations is the slit-stepping strategy illustrated in Figure~\ref{fig:multiplex}. We observed the same MSA mask configuration in a grid of 63 different pointings: 9 steps along the dispersion direction at a width of a slitlet (0\farcs2), with 7 dithers along the cross-dispersion direction stepped by the width of a barshadow (0\farcs075). A single exposure of 19.5 minutes was taken at each of the 63 dithered positions, using the NRSIRS2 readout mode with 16 groups. 
This 9$\times$7 dither pattern provides nearly uniform depth, where each sky position is observed with six exposures on-source (117 minutes) and one exposure affected by a bar shadow. 
Each target galaxy is covered with a minimum of 3 slitlets of $\sim$0\farcs5 length each. We add up to 2 additional slitlets wherever possible, avoiding MSA slit collisions. 
This results in data cubes spanning 1\farcs8$\times$(2\farcs0--3\farcs0) depending on the number of slitlets (3--5), as illustrated in Figures~\ref{fig:color_image} and \ref{fig:multiplex}. The region with uniform full exposure coverage is 1\farcs8$\times$(1\farcs1--2\farcs1), corresponding to approximately 15$\times$(10-18) kpc at our target redshifts. 
We emphasize that the slit-stepping strategy is designed for \textit{contiguous uniform depth} in this field of view. In contrast, a rectangular grid of slitlets (e.g., 9$\times$5) on the MSA has only $\sim$65\% illumination with approximately one-third of the area obscured by bar shadows. Our observations effectively remove these bar shadows to provide contiguous sky coverage Figure~\ref{fig:multiplex}).

The areal coverage set by the number of slit-steps (9) and slitlets per target ($\geq 3$) is chosen to sample galaxy radii $\gtrsim 3 R_s$, where $R_s \simeq 2.5$~kpc is the typical scale radius of star-forming galaxies at the upper mass range of our sample ($M_* \sim 10^{11}~\Msun$). A key goal is to map galaxy kinematics beyond the ``turnover radius'' ($R > 2.2 R_s$) corresponding to the maximum circular velocity for an exponential disk \citep{freeman1970}. 
The depth of $\sim$2 hours effective exposure time is sufficient to reach $\sim 5\sigma$ detections of \Ha\ emission in an 0\farcs1$\times$0\farcs2 resolution element at $R \simeq 3 R_s$ for typical $z\simeq1$ galaxies across the targeted mass range, based on surface brightness profiles measured from HST grism spectroscopy \citep{nelson2016}. 
In addition, the total program time of $\sim$29 hours including overheads allowed all spectra to be taken in a single visit, to avoid positional uncertainty which may arise from multiple visits.
The program is thus optimized to map \Ha\ emission in $z\sim1$ galaxies out to several disk scale radii, with simultaneous coverage of additional strong optical emission lines. This enables our scientific goals to map gas kinematics, metallicity, ionization mechanisms, dust attenuation, and star formation rate on kpc scales.

\subsection{Comparison of slit-stepping vs. IFU modes}
\label{sec:IFU_comparison}

Here we briefly compare our slit-stepping strategy with the conventional integral field unit spectrograph mode of JWST/NIRSpec. 
The IFU is effectively limited to observing a single galaxy at a time given its 3\arcsec$\times$3\arcsec\ field of view. The IFU is also less sensitive, requiring approximately 4$\times$ longer on-source integration than the MSA to reach the same sensitivity in a fixed aperture (e.g., 0\farcs1$\times$0\farcs2). Thus the IFU would require $\sim$8 hours on-source per galaxy to reach the same depth as our MSA observations. 
In contrast we obtained spatially resolved spectroscopy for 43 galaxies with only $\sim$20 hours of exposure time, thanks to the MSA's multiplexing capability. This represents a factor $\sim 15\times$ advantage in exposure time alone; the total time advantage may be up to two times greater considering larger total slew time overheads to observe a sample of objects with the IFU mode. 

The primary disadvantage of the MSA for this work is the 0\farcs2 slitlet width, which undersamples the PSF and is coarser than the IFU's 0\farcs1 spaxels. The MSA nonetheless provides sufficient spatial sampling for our objectives. 
Another disadvantage of the MSA is that the spectral coverage varies depending on target position. This is a modest effect in our case: 12\% of targets (5/43) do not have \Ha\ coverage due to their location on the MSA mask. The spectral coverage effect is most severe with the high resolution gratings (including G140H used here); a program using medium or low resolution gratings would be less affected. 
Ultimately the most significant of the above effects are the MSA's sensitivity and multiplexing advantages, which deliver an order of magnitude faster survey speed for building spatially resolved spectroscopic samples. 
In sum, MSA slit-stepping is strongly preferred for our scientific program due to large efficiency gains compared to the IFU mode. 

In general, the slit-stepping mode is powerful for surveys making use of the MSA's multiplexing capability, whereas the IFU mode is more effective for targets which are sparse on the sky. Here we roughly quantify the relative efficiency of the two modes for a general use case. Assuming that $N_{mask}$ targets can be observed on a single MSA mask, with spatial coverage required over a diameter $D$ (such that the number of slit-steps required is $D$/0\farcs2), the efficiency gain in exposure time required for MSA slit-stepping compared to the IFU is
\begin{equation}
\label{eq:efficiency_gain}
    G = \frac{\mathrm{IFU~exposure~time}}{\mathrm{slit~stepping~exposure~time}} \approx 3.5 N_{mask} \left( \frac{0\farcs2}{D} \right).
\end{equation}
The factor of $\approx 3.5$ arises from the IFU's lower sensitivity (contributing $\sim 4\times$) multiplied by the fractional area covered by MSA slitlets ($\sim 0.87 =$ 0\farcs46 slitlet length out of 0\farcs53 pitch, with 0\farcs075 obscured by a bar shadow). 
Equation~\ref{eq:efficiency_gain} is valid for $D \leq 3\arcsec$ such that targets can be observed with a single IFU pointing. For more extended sources requiring $N_{IFU}$ pointings per target, the result becomes
\begin{equation}
\label{eq:efficiency_gain_v2}
    G \approx 3.5 N_{mask} \left( \frac{0\farcs2}{D} \right) N_{IFU}.
\end{equation}
A gain $G>1$ indicates the slit-stepping approach is more efficient. As an example, for cases requiring a 3\farcs0 diameter field of view, slit-stepping is more efficient when $\gtrsim5$ targets can be observed on an MSA mask. The IFU is more efficient ($G<1$) if targets are more sparse on the sky. The slit-stepping mode can also be more efficient for observing individual sources which are highly extended ($D \gtrsim 10\arcsec$, where $N_{IFU}$ becomes large). 
We note that individual science cases may prefer the IFU for reasons such as spatial sampling or other complexities. Additionally some cases require a modest sample size $N_{sample} < N_{mask}$, such that $N_{mask}$ should be replaced by $N_{sample}$ for purposes of total efficiency. 
Otherwise, for use cases in which MSA spectra are suitable, Equations~\ref{eq:efficiency_gain} and \ref{eq:efficiency_gain_v2} offer approximate guidance on which mode is more appropriate in terms of survey speed.

% Quantify when slit-stepping vs IFU is more efficient: efficiency factor \propto (number of objects per MSA mask) / (required field of view per object, i.e. number of steps)

  % Although JWST NIRSpec features a conventional integral field unit spectrograph, this mode is limited to observing only one galaxy at a time. In comparison, NIRSpec's MOS mode possesses multiplexing capability, combined with its nearly 2x higher sensitivity.  Benefiting from MOS sensitivity to enhance observing time efficiency, we simultaneously sample all targets employing the slit-stepping methodology. Our science objectives necessitate the ability to distinguish a typical intrinsic velocity dispersion for thin disks ($\sigma$ < 30km/s), which requires an \Ha\ SNR = 5 at 3Rs per spatial element. Planning with this constraint, the cumulative exposure time for the slit-stepping methodology in this configuration (9x7) amounts to $\sim$21hrs, leading to a superb time efficiency of 30min/galaxy. To attain equivalent sensitivity (per 0.1x0.2 resolution element) the IFU would require approximately 8hrs of on source integration.  Despite IFU’s ability for slightly better spatial sampling (0.1” x 0.1” vs 0.1” x 0.2”), achieving the desired sensitivity on this finer scale would demand substantially more integration time. Comparing 8hrs to 30mins, slit-stepping showcases its power with over 15x faster exposure time efficiency needed to build extensive IFS datasets. Considering overheads, MSA slit-stepping amounts to $\sim$29hrs, making it up to 30x faster method than IFU.

\subsection{Reduction Pipeline and Data Cube Construction}
\label{sec:pipeline}

As our slit-stepping observing methodology is non-standard, we have developed a custom data reduction procedure to process the raw data and ultimately construct data cubes for each target. 
We primarily use procedures from the JWST Science Calibration Pipeline developed by the Space Telescope Science Institute (STScI) and augment this pipeline with custom steps to improve performance with the small sub-slitlet dithers. We use versions 1.9.6 (CRDS file ``jwst\_1075.pmap'') and 1.10.2 (CRDS file ``jwst\_1105.pmap'') of the standard reduction pipeline for stages 1 and (2, 3) respectively. We selected version 1.9.6 for Stage 1 to avoid the significant negative pixel artifacts that occurred with version 1.10.2. Crucially, we use a new pseudo-IFU cube building class in post-processing, as this observing strategy is not supported by the default STScI pipeline. The steps are described below.

\begin{enumerate}
    \item Stage 1\\
    %The JWST Science Calibration Pipeline consists of 3 processing stages. Stage 1 (the ``calwebb\_detector1'' module) applies detector-level corrections for all groups, transforming the data into usable slope images. 
    Stage 1 (out of 3) of the JWST Science Calibration Pipeline (the ``calwebb\_detector1'' module) applies detector-level corrections for all groups, transforming the data into usable slope images.
    %This stage is designed to process uncalibrated raw data from all instruments and modes. 
    %\textcolor{magenta}{In this stage we use version 1.9.6 of the reduction pipeline with CRDS file ``jwst\_1075.pmap''.} 
    To improve the identification and treatment of artifacts (e.g., snowballs and other cosmic ray traces; \citealt{green2020_snowballs}), we modify the default values of the following parameters in the jump detection step:

    \noindent \textit{expand\_large\_events}=True, \\
    \textit{sat\_required\_snowball} = False,\\ 
    \textit{min\_jump\_area} = 10.

    A default value for the parameter controlling the expansion of the number of pixels flagged around large cosmic ray events: snowballs, %cosmic events 
    is kept ‘True’ ({\textit{expand\_large\_events}}). 
    To enhance data processing efficiency, JWST's onboard hardware averages or drops frames to create groups. This can lead to the snowball detection algorithm missing saturated cores occurring between frames within a group, resulting in delays in accurate snowball detection (\href{https://www.stsci.edu/files/live/sites/www/files/home/jwst/documentation/technical-documents/_documents/JWST-STScI-008545.pdf}{section 3.5 in JWST technical report}).
    To address this issue, we set {\textit{sat\_required\_snowball}} to ‘False’ (default = ‘True’) and increase the default minimum number of pixels ({\textit{min\_jump\_area}} = 5) to 10. This bypasses the need for presence of saturated pixels within the defined jump circle and triggers the algorithm expanding the area around snowballs.
%    However, the default minimum number of pixels ({\textit{min\_jump\_area}}) required to trigger the flagging of the presence of \textcolor{magenta}{snowballs} %cosmic ray events 
%    was too small (= 5), the result of which was that majority of \textcolor{magenta}{snowballs} %cosmic ray events 
%    were not detected.  Thus, we increase the number of required pixels to 10. We set {\textit{sat\_required\_snowball}} to ‘False’ (default = True) to bypass the requirement for the presence of saturated pixels within the defined jump radius in order to trigger the artifact treatment. 
    The combination of these parameters leads to a successful detection and treatment of snowball events.
    We refer readers to \href{https://www.stsci.edu/files/live/sites/www/files/home/jwst/documentation/technical-documents/_documents/JWST-STScI-008545.pdf}{this JWST technical report} for a detailed discussion.
    \\
    The output is a corrected 2D count rate slope 
    % (“\_rate.fits”) 
    image in units of counts/s. The count rate files are used as input for remaining stages of the pipeline.

    \item Pre-processing: $1/f$ noise correction \\
    While the NRSIRS2 readout mode is designed to mitigate $1/f$ read noise in the NIRSpec detectors, some correlated noise remains in the form of vertical stripes and edge effects. %This is due to reliance on reference pixels with an assumption that the readout noise is covariance stationary, whereas thermal instability causes non-stationary covariance noise. 
    We use the NSClean algorithm with its default settings %and manually designed masks 
    \citep[as detailed in ][]{rauscher2023_nsclean} to remove the residual noise. % from the count rate file, 
    In the manual mask design, \cite{rauscher2023_nsclean} selects all spectral traces and illuminated pixels, then inverts the selection to create a background model. Background pixels are weighted to address uneven sampling, with fewer unilluminated pixels between spectral traces and greater prevalence at the top/middle/bottom of each count rate file.
    This produces output count rate files with uniform and nearly complete removal of correlated noise  (\href{https://jwst-pipeline.readthedocs.io/en/stable/jwst/nsclean/main.html}{see JWST documentation}).
    
     % The amount of noise that persists, in short, is due to how IRS2 mode operates. Built on principal component analysis, IRS2 relies on reference pixels to remove the noise and assumes that the read out noise is covariance stationary. In reality, thermal instability causes non stationary covariance noise. 

\begin{figure*}
    \centering
    \includegraphics[width=\linewidth]{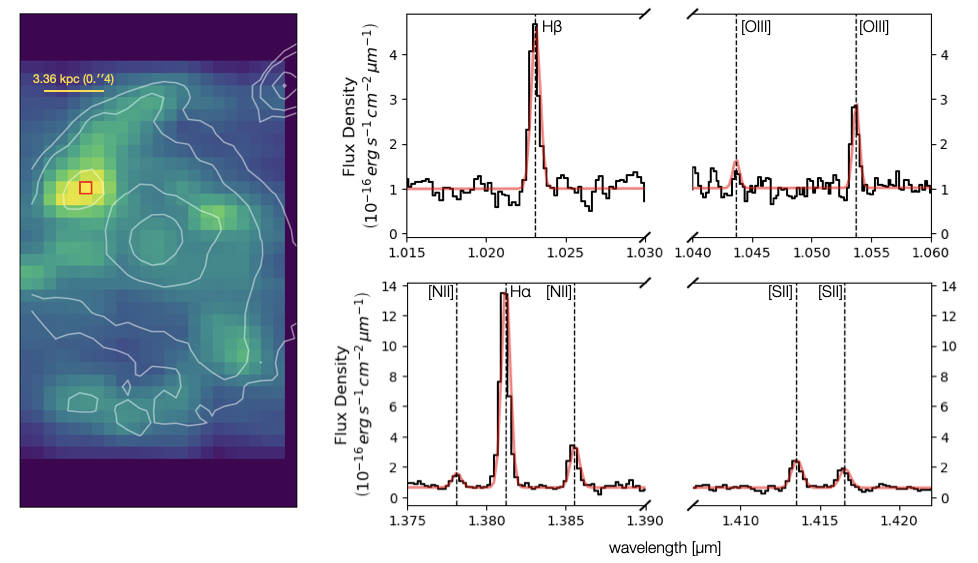}
    \caption{The left panel features a collapsed 2-dimensional image from the data cube centered around the \Ha\ line at rest-frame $\sim$6563~\AA. Overlaid white contours indicate the galaxy continuum morphology observed in JWST/NIRCAM F444W imaging. 
    We note that the bright continuum source in the upper-right corner is not associated with the main spiral galaxy; our slit-stepping data spectroscopically confirm it at a slightly lower redshift $z=1.066$. 
    The right panels present the 1-dimensional spectrum of a single bright spaxel (shown with the red square at left), zoomed in around the key diagnostic emission lines: \Hb, \Oiii, \Nii, \Ha, and \Sii. Fits to the emission lines are shown in red and are described in the text. All key lines are well detected. Spatial mapping of these emission lines enables characterization of the kinematics, gas-phase metallicity and radial gradients, dust attenuation, star formation rate profile, and nuclear activity in this galaxy. We note that this is only one example of the 43 galaxies observed with our slit-stepping program.}
    \label{fig:int_spec}
\end{figure*}

    \item Stage 2\\
    Stage 2 (``calwebb\_spec2'') of JWST’s Science Calibration Pipeline is dedicated to spectroscopic data refinement. Stage 2 applies further instrument-level corrections to slope images, generating calibrated exposures. 
     % (“\_cal”). 
    Our customized processing includes the following steps from this stage: WCS assignment, MSA flagging, extraction of 2D arrays from spectral images, flat field and barshadow correction, flux calibration, and resampling. We do not include the standard pathloss correction step, and instead perform it in post-processing, which we explain in more details in Step 4.iii.
    %\textbf{Pathloss correction in the standard reduction pipeline currently performs only for a standard number of slit-lets (= 3)\footnote{Solution expected in February 2024; \href{https://jwst-docs.stsci.edu/jwst-calibration-pipeline-caveats/known-issues-with-jwst-data-products}{JDox known issues}}. As our program often uses $>$ 3 slit-lets on a given target, we are not able to include pathloss correction at this stage, and perform it in post-processing (see Step 4).}  
    %Stage 3 produces final rectified 2D spectra
     % (“\_s2d”) 
    %for each target from the calibrated exposures; in this stage we apply only the resampling step. 

    \item Stage 3 and post-processing\\
    i. In Stage 3 (``calwebb\_spec3'') we perform only the resampling step to create final rectified and combined 2D spectra for each target from calibrated exposures. %Each 2D spectrum is resampled using WCS and distortion data, and the resampled spectra from individual detectors are combined into a single undistorted spectrum for each target.
    This step involves taking 2D spectra from each individual detector, resampling them using WCS and distortion information, and combining them into a single undistorted 2D spectrum for each target.
    
    ii. A number of outliers, including cosmic rays and other artifacts (e.g., hot pixels) are apparent in the processed 2D spectra, the treatment of which remains an open issue in the standard pipeline. To remove the outliers we apply the astro-SCRAPPY\footnote{astro-SCRAPPY: \href{https://zenodo.org/record/1482019}{https://zenodo.org/record/1482019}} cosmic ray cleaning code, which is based on the L.A.Cosmic algorithm \citep{vanDokkum2001_lacosmic}. 
     % We apply astro-SCRAPPY (author: Curtis McCully), which is built on Pieter van Dokkum’s L.A. Cosmic algorithm, designed for the removal of cosmic ray artifacts, is based on a form of Laplacian edge detection — the algorithm detects cosmic rays by the sharpness of their edges. 

    iii. The role of the pathloss correction function is to account for geometrical and diffraction losses of light. Pathloss correction in the standard pipeline relies on theoretical optical models to correct for pathloss effects for non-centered point sources and those uniformly filling the slit \citep{ferruit2022}.
    The current implementation of pathloss correction in the STScI reduction pipeline performs only for a standard number of slitlets (one and three). %\footnote{Solution expected in February 2024; \href{https://jwst-docs.stsci.edu/jwst-calibration-pipeline-caveats/known-issues-with-jwst-data-products}{JDox known issues}}. 
    As the targets in our program are often observed with $>$3 slitlets, we perform this correction outside of the standard pipeline.
    We assume a uniformly illuminated slit to correct pathloss effects for extended sources in our slit-stepping program \citep[][]{nanayakkara2023}. We extract a 1D pathloss function from the pipeline's uniform pathloss reference file, and correct the 2D spectra of all exposures considering the relevant wavelength range for each target (due to the pathloss model wavelength dependence).

    \item Cube construction\\
    Our slit stepping pattern, as described in Section~\ref{sec:observing_strategy}, consists of 9 dispersion ``steps'' (0\farcs2 each) and 7 cross-dispersion dithers (0\farcs075 each). Our cube building process combines the 63 individual pointings into a 3D data cube for each target. We first combine 2D spectra in the cross dispersion direction to produce a single 2D spectral slice at each step. We resample individual 2D spectra from each of the 7 dithered exposures (at each dispersion step) onto a common grid with finer spatial pixel scale, such that original pixels are Nyquist sampled. We then take the median of all exposures (with 7 exposures in the central regions, and fewer in the outer 0\farcs45). % and rebin to the original 2D pixel scale. 
    A median combination is used to minimize residual effects from cosmic rays, bar shadows, alternating column noise, and other artifacts. This is sufficient for the results presented herein, and this key processing step can be updated with an outlier-rejected weighted mean or other optimized algorithm following future improvements to the earlier pipeline stages. 
    The combination of individual steps is straightforward: we append the 2D spectra at each of the 9 steps into a 3D data cube, with 9 spaxels of 0\farcs2 in the dispersion direction. 
    The Nyquist sampled cross-dispersion spectra can be rebinned to the original pixel scale, resulting in 0\farcs1$\times$0\farcs2 spaxels. For analysis and display purposes in this paper, we instead interpolate the data to square spaxels of 0\farcs08$\times$0\farcs08. This spaxel size corresponds to 0.7 kpc at redshifts $z=1$--1.7. 
    
\end{enumerate}

\begin{figure*}
    \centering
    \includegraphics[width=\linewidth]{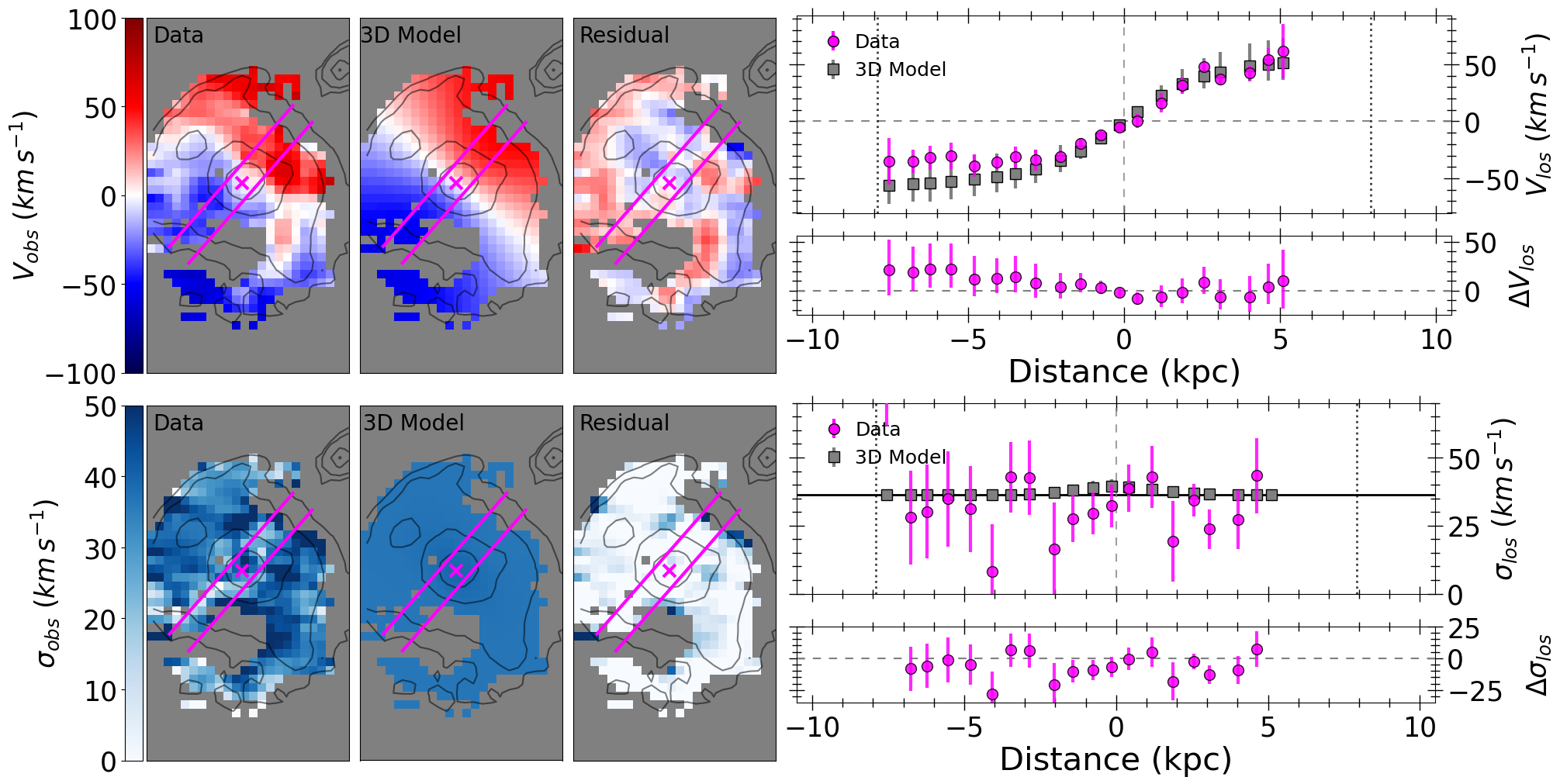}
    \caption{Kinematic measurements and best-fit disk model of the target galaxy. The \textit{left panels} show: velocity $V_{obs}$ and velocity dispersion $\sigma_{obs}$ measured from a joint fit to \Ha\ and surrounding lines in each spaxel (\textit{left}), the corresponding $V$ and $\sigma$ maps from our best-fit disk model (\textit{center}) and the residuals (\textit{right}). The kinematic center of the disk model is marked with a magenta 'x', and is located near the central continuum flux density peak in the F444W filter (shown with contours in each panel). Magenta lines show the location of a pseudo-slit along the kinematic major axis, from which we extract the 1-dimensional rotation curve and velocity dispersion (\textit{right panels}). For the 1-D curves we similarly show the measured data, best-fit model, and residuals ($\Delta V$ and $\Delta_{\sigma}$). 
    % We additionally show the model's intrinsic rotation curve corrected for inclination as a black line with shading representing the formal 1$\sigma$ uncertainty. 
    Overall the disk model provides a good fit to the observed data, supporting that the nebular emission arises from a rotating disk viewed nearly face-on. The model residuals are $\lesssim 20$~\kms\ and comparable to the measurement uncertainties, as can be seen in the right panels. The high angular resolution and face-on nature of this system provide an excellent measurement of the local velocity dispersion $\sigma$.
    } % Change later
    \label{fig:Ha_maps}
\end{figure*}

\subsection{Emission Line Fitting}
\label{sec:emission}

The results in this paper are based on spatially resolved measurements of the strong rest-frame optical nebular emission lines. The angular resolution is limited by NIRSpec/MSA which undersamples JWST's point spread function (PSF), especially in the dispersion direction with 0\farcs2 slitlet width. 
% [Kris said there was no smoothing in the end]
%For display purposes and to increase the signal-to-noise (S/N), we spatially smooth the data to produce an approximately circular PSF of 0\farcs26 FWHM \textcolor{red}{\textbf{[UPDATE! It was actually done with less smoothing, no?]}}. This is accomplished by convolving the datacubes with an elliptical Gaussian kernel \textcolor{red{[of \$DIMENSIONS -- or is it smoothed only along the cross-dispersion axis?]}. 
For the galaxy analyzed in this work, we achieve S/N~$>5$ in the \Ha\ emission line for individual spaxels spanning $\sim$2\farcs5, with all additional targeted lines (\Nii, \Sii, \Hb, \Oiii) well-detected in bright individual spaxels. The \Ha\ narrowband flux map and example emission line detections from a single spaxel are shown in Figure~\ref{fig:int_spec}. 

We fit the emission lines within each spaxel using Gaussian profiles, which provide a good fit. The \Ha, \Nii~$\lambda\lambda6548,6585$, and \Sii~$\lambda \lambda6717,6732$ lines are jointly fit with a five-component model plus a linear continuum. All lines are assumed to have the same velocity ($V$) and velocity dispersion ($\sigma$). The flux of each line is a free parameter, except for the \Nii\ doublet whose flux ratio is fixed to the theoretical value of 2.942. The \Hb\ and \Oiii~$\lambda\lambda4960,5008$ lines are similarly fit with a three-component model with the \Oiii\ doublet flux ratio fixed to 2.984. 
Example maps of the \Ha\ flux, velocity, and velocity dispersion are show in Figure \ref{fig:Ha_maps}. The velocity dispersion $\sigma_{obs}$ is corrected for NIRSpec's spectral resolution ($R\sim2700$ for the G140H grating) by subtracting the line spread function FWHM in quadrature from the best-fit Gaussian FWHMs. We obtained the instrumental FWHM~$\simeq5.2$~\AA\ by interpolating the G140H resolution curve across the wavelength range of emission lines used here. 
In this work we adopt a systemic redshift $z_{sys}=1.10417\pm0.00018$ corresponding to the central region of the galaxy. Velocities in Figure~\ref{fig:Ha_maps} are shown relative to this redshift.

%\begin{figure*}
%    \centering
%    \includegraphics[width=0.75\linewidth]{flux_vel_maps_may16.png}
%    \caption{\Ha\ flux, velocity, and dispersion maps}
%    \label{fig:Ha_maps}
%\end{figure*}

\section{Physical Properties} 
\label{sec:results}

\subsection{Disk Kinematics}
\label{sec:kinematics}

We derive kinematic properties using GalPaK \citep[][]{bouche2015}, a fitting software optimized for IFU data. We fit the observed \Ha\ emission in the example target galaxy via forward modeling of galactic disk models, accounting for the line-spread function (LSF) and point-spread function (PSF). 
The LSF is assumed to be a Gaussian of FWHM~$\approx5.2$~\AA\ based on NIRSpec documentation, and the PSF is estimated as an elliptical Gaussian with FWHM~$=0\farcs08 \times 0\farcs2$, accounting for undersampling by the 0\farcs2 MSA slitlets.
% GalPaK3D has been shown to work well with VLT/MUSE data \citep[][]{bacon2015, contini2016} as well as KMOS data \citep[][]{mason2017}. The full description of the fitting procedure is referenced in \cite{bouche2015}; however, we will discuss the key modeling processes pertaining to this paper.
In this paper we assume a rotating disk geometry with a thickness of scale height $h_z$. 
The rotation curve is modeled as an arctangent function \citep[][]{courteau1997}:
\begin{equation}
    V(r) = \frac{2}{\pi}V_{max}\sin(i)\arctan\left(\frac{r}{r_t}\right)
\end{equation}
where $r$ is the distance along the major axis of the galaxy in the plane of the sky, $V_{max}$ is the asymptotic velocity in the plane of the disk, $i$ is the inclination of the disk, and $r_t$ is the turnover radius. We also model the total line-of-sight velocity dispersion, $\sigma_{tot}$, which in general has three components added in quadrature: a local isotropic velocity dispersion $\sigma_d$ from the self-gravity of the disk, which is given by $\sigma_d(r) = h_z V(r)/r$ for a compact thick disk; a mixing term $\sigma_m$ due to the mixing of velocities along the line of sight in a disk with non-zero thickness; and an intrinsic velocity dispersion $\sigma_0$ (assumed to be isotropic and spatially constant), which represents the dynamical ``hotness'' of the disk.

Given the rich spiral morphology of \Ha\ emission, which differs somewhat from the broad-band continuum (Figure~\ref{fig:int_spec}), we do not model the full 3D flux distribution in the data cube. Instead we fit the velocity and dispersion maps measured from \Ha\ and the surrounding metal lines as discussed in Section~\ref{sec:emission}. 
We use GalPaK to construct model 2D maps of the velocity and dispersion, and perform a Markov chain Monte Carlo (MCMC) parameter space exploration with the Python software emcee \citep{foreman-mackey2013_emcee}. A Gaussian prior is adopted for the center of the galaxy, using the position of peak stellar continuum flux in the NIRCam F444W image. 
The F444W filter corresponds to rest-frame K band ($\sim 2.1$~\um) which is expected to be a good tracer of the stellar mass distribution \citep[e.g.,][]{bell2001}.
For all other parameters we adopt a flat prior (including, e.g., $V_{max}$ and $\sigma_0$).
The likelihood function for our model is defined by agreement with the measured velocity $V$ and dispersion $\sigma$ maps: 

\begin{equation}
    \log \mathcal{L} \sim -0.5 \times \sum_{X \in \{V,\sigma\}}\sum_{j}\left(\frac{X_{obs,j}-X_{model,j}}{Err_{X,j}}\right)^2
\end{equation}

which sums over all spaxels ($i$) in both $V$ and $\sigma$. Here $X_{obs}$, $X_{model}$, and $Err_{X}$ are the 2D maps of the data, extracted model, and standard deviation measurement uncertainty, respectively. 
The best-fit $V_{model}$ and $\sigma_{model}$ maps along with the residuals ($X_{obs} - X_{model}$) are shown in Figure~\ref{fig:Ha_maps}. We also show 1D rotation and velocity disperion curves extracted along the best-fit major axis. 

Overall the disk model provides a good fit to the kinematic measurements. The residuals are $\lesssim20$~\kms\ for both $V$ and $\sigma$. The best-fit inclination $i = 12.9^{+2.7}_{-1.1}$ degrees is nearly face on, which is also clear from the imaging data. We note that the source morphology was not used to constrain the fit, such that the model provides an independent confirmation of the face-on disk geometry. 
While the line-of-sight rotational velocity $V_{max} \sin(i)$ is tightly constrained by the data, the large inclination correction factor (i.e., $\frac{1}{\sin(i)} \simeq 4.5$) makes $V_{max} = 290 \pm 40$~\kms and $V/\sigma = 8.1\pm1.5$ relatively uncertain. 
 % This is apparent in the rotation curve in Figure~\ref{fig:Ha_maps}. 
As such we refrain from detailed analysis of the rotation curve and dynamical mass profile in this galaxy. We note that the $V/\sigma$ value is characteristic of well-settled disks, and agrees well with predictions from the TNG50 simulation at this redshift and stellar mass \citep{pillepich2019}.
Such high values of $V/\sigma$ are also found in the FIRE zoom-in simulations after disk settling \citep[][]{stern2021, gurvich2023}.
The majority of our targets have a higher (more edge-on) inclination which allows better constraints on the circular rotation velocity.
%\textcolor{orange}{the velocity dispersion $\sigma = 36.18^{+4.60}_{-4.11}$~\kms\ } 
On the other hand, the velocity dispersion $\sigma_0 = 36^{+5}_{-4}$~\kms\ is determined with good precision, and this is the key quantity which describes the dynamical state of the disk (e.g., cold thin disk vs. turbulent thick disk). 
The \Hb\ and \Oiii\ emission lines give consistent results.
In this case the $\sigma$ value observed from a nearly face-on inclination corresponds closely with the vertical component of velocity dispersion, which sets the disk scale height.

\begin{figure}
    \centering
    \includegraphics[width=\linewidth]{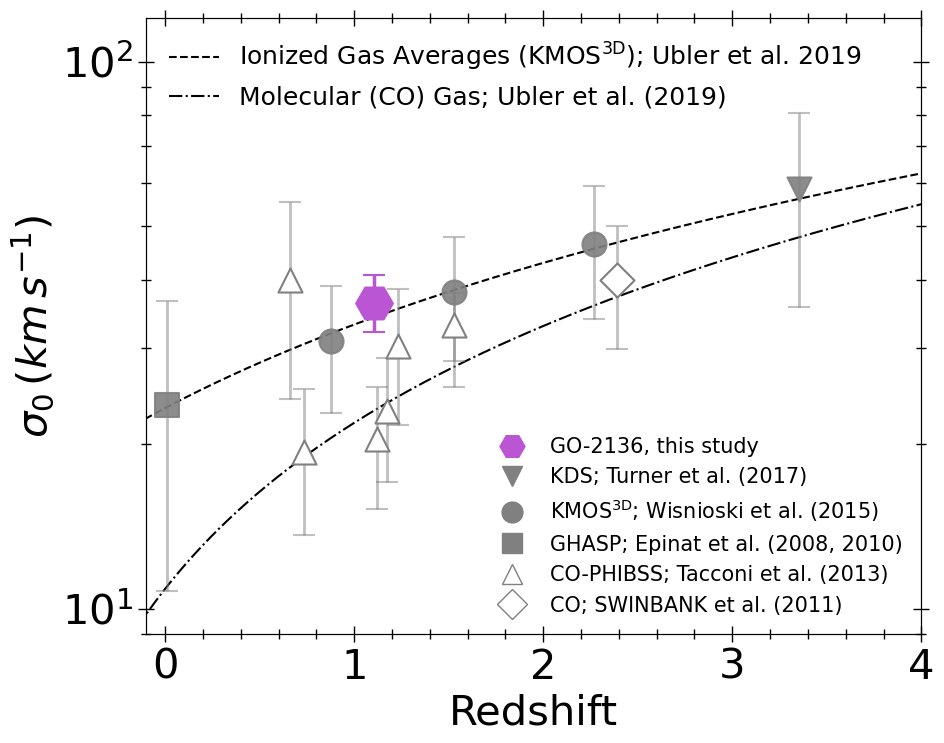}
    \caption{Velocity dispersion measured for our galaxy (violet hexagon) compared to previous observational results spanning redshifts up to $z\lesssim4$ (KDS: \citealt{turner2017}; KMOS-3D: \citealt{wisnioski2015}; GHASP: \citealt{epinat2008, epinat2009}; PHIBBS: \citealt{tacconi2013}; \citealt{swinbank2011}). We show the approximate redshift evolution separately for ionized gas such as \Ha\ (filled symbols) and molecular gas traced by CO (open symbols). We show the average redshift evolution curves for ionized (dashed) and molecular gas (dash-dotted) from \cite{ubler2019}. 
    Our measurement from JWST slit-stepping data aligns well with the average ionized gas measurements at $z\sim1$.}
    \label{fig:sig0_z_plot}
\end{figure}

\begin{figure*}
    \centering
    \includegraphics[width=\textwidth]{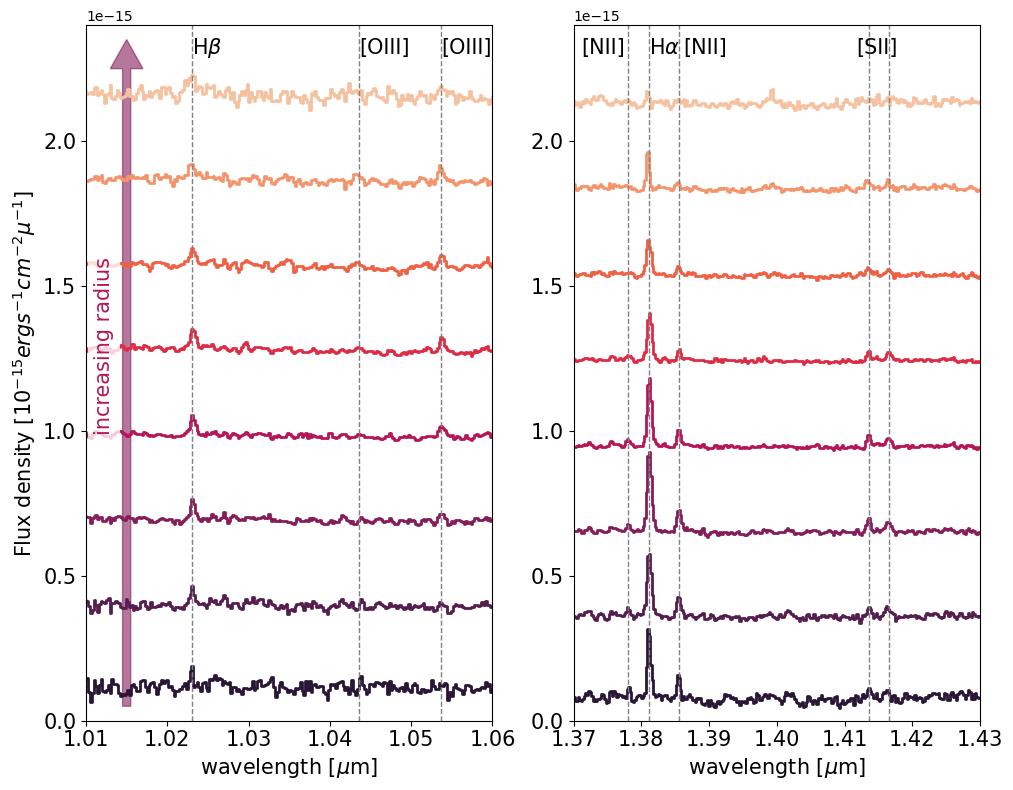} 
    \caption{Average flux density per spaxel in each radial bin. Successive bins are offset vertically for display purposes, extending from the galaxy's center (bottom spectra) to the outermost bin at $\sim$13~kpc (top-most spectra), as indicated by the arrow at left. The bins are constructed from deprojected galactocentric radius as described in the text, with a bin width of 1.7~kpc (0\farcs2). 
    The outermost bin is not used in our analysis due to low S/N of the emission lines.
    Vertical gray dashed lines highlight the main diagnostic emission lines of \Hb, \Oiii, \Nii, \Ha, and \Sii. We observe a clear radial trend in emission line ratios, with \Nii/\Ha\ decreasing and \Oiii/\Hb\ increasing at larger radii. This is explained by a smooth gas-phase metallicity gradient as described in Section~\ref{sec:metallicity}.}
    \label{fig:radial_flux}
\end{figure*}

In Figure \ref{fig:sig0_z_plot} we compare $\sigma_0$ for the galaxy studied here with average values compiled from various surveys as a function of redshift. 
Despite the clear spiral disk morphology, the measured $\sigma_0$ is higher than for typical spiral galaxies at $z\simeq0$, and comparable to the largest values found in the GHASP sample \citep{epinat2008}. 
% Mean \sigma in the GHASP sample is 24\pm5 km/s, from Epinat+2008. 
However we find excellent agreement with the average trend seen in KMOS$^\mathrm{3D}$ \citep{wisnioski2015} and other kinematic surveys of ionized gas \citep[e.g.,][]{simons2017}. While this is only a single object whose properties may not be representative, it appears to support the picture of velocity dispersion increasing with redshift. Our best-fit $\sigma_0$ may slightly underestimate the true value due to possible slit illumination effects. Our analysis assumes uniform slit illumination, whereas a partial illumination (e.g., with bright nebular emission concentrated on one side of the 0\farcs2 slit) would cause the line widths (e.g., $\sigma$) to appear smaller. 
Such an effect could cause variation in both the measured $\sigma$ and $V$, along with larger scatter in line widths potentially reaching values smaller than the assumed instrument resolution. We would also expect the velocity dispersion measured from \Hb\ to be systematically smaller than \Ha\ in the presence of non-uniform slit illumination (due to the smaller PSF at shorter wavelengths). Instead we find relatively small scatter in velocity dispersion, with values from \Hb\ and \Oiii\ being slightly larger on average (by $13\pm11$\%) than those measured from \Ha\ in radial bins. We thus conclude that the bias from non-uniform slit illumination is comparable or smaller than the measurement uncertainty. 
Future efforts may be able to better quantify the effects of non-uniform slit illumination by forward-modeling the light distribution (e.g., based on imaging or grism spectroscopy).
 %Such an effect could cause the observed $\sigma$ to vary by a factor of $\sim 2\times$, with values possibly reaching half the nominal line spread function resolution. The measured $V$ would also vary depending on the illumination profile. However we observe a relatively small scatter in $\sigma$ across the galaxy, with no values significantly below the instrument resolution, which together suggest that this effect is minor. 

In summary we find that the dynamical state of the disk, as indicated by local velocity dispersion, is more turbulent than in $z\sim0$ spirals and comparable to that inferred from large ground-based samples at $z\sim1$ \citep[e.g.,][]{wisnioski2015,stott2016}. 
In contrast to typical ground-based data, our measurement of $\sigma_0$ is especially robust to beam smearing effects, thanks to the nearly face-on inclination and the exquisite space-based point-spread function of the data. 
The best-fit kinematic major axis (PA~$=139$ degrees) and disk inclination ($i = 12.9$ degrees) are used to construct deprojected radial bins in the following sections.

\begin{figure}
    \centering
    \includegraphics[width=\linewidth]{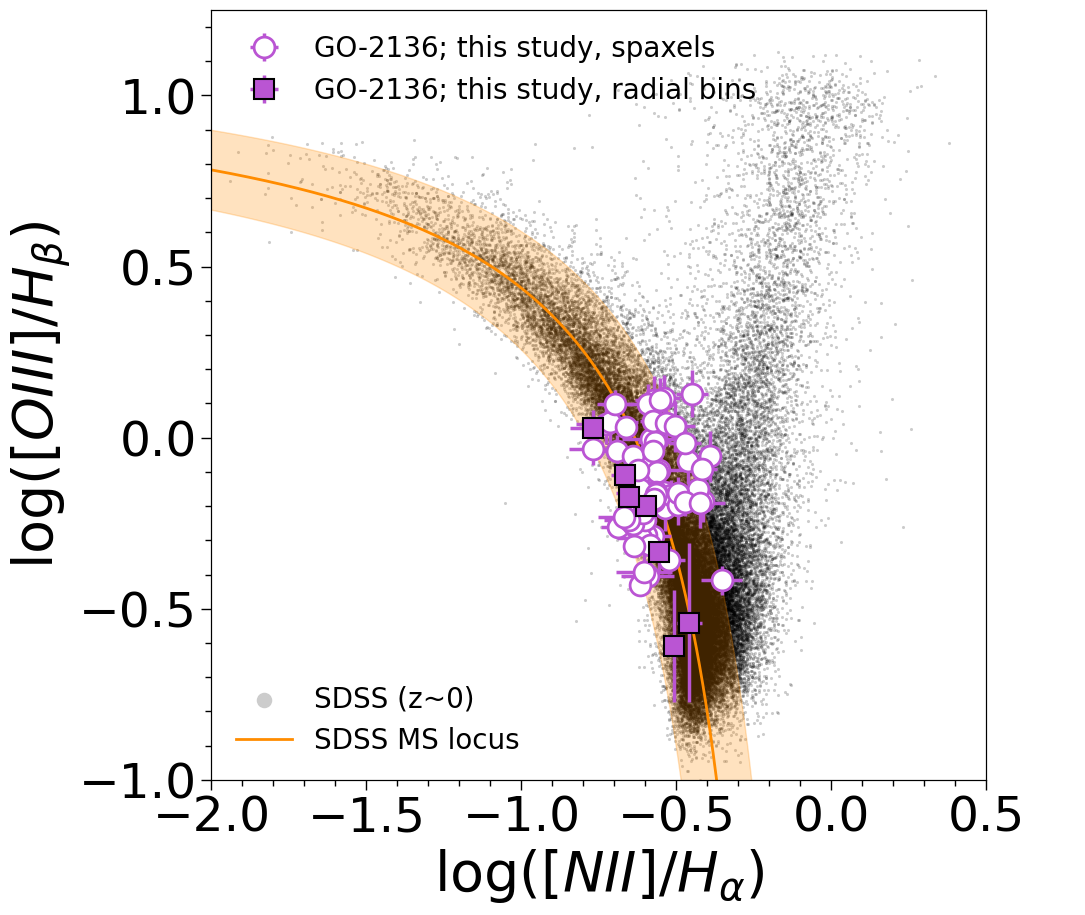}
    \caption{BPT diagram of the \Oiii/\Hb\ versus \Nii/\Ha\ flux ratios, with measurements for both individual spaxels (open circles) and radial bins (filled squares). 
    These are overlaid on a large sample of $z\sim0$ galaxies from the Sloan Digital Sky Survey (SDSS; \citealt{abazajian2009_sdss_dr7}). The star-forming locus described by \cite{kewley2013} is shown with the orange line and shading, while the AGN sequence extends to the upper right. 
    Both the individual spaxels and radial bins for our target galaxy consistently lie along the locus of star-forming galaxies, with radial trends indicative of metallicity gradients (see Section~\ref{sec:metallicity}). 
    Based on this BPT diagnostic, we do not detect any signature of an AGN in this target; nebular emission is dominated by star forming regions even within the central resolution element.}
    \label{fig:BPT_diagram}
\end{figure}

\subsection{Flux Ratios and BPT Diagram}\
\label{sec:BPT_diagram}

To classify the dominant energy source of line emission within this galaxy and to establish the degree of nuclear supermassive black hole activity, we examined the Baldwin–Phillips–Terlevich (BPT) diagram \citep[][]{baldwin1981} of \Oiii/\Hb\ versus \Nii/\Ha\ flux ratios. Given the moderate S/N of \Oiii\ and \Hb\ emission within typical individual spaxels, we use radial annular bins. The bins are constructed using the deprojected radius based on the best-fit inclination and position angle derived from our kinematic disk model (Section~\ref{sec:kinematics}). 
%the average radii between each consecutive bin varies radially by $\sim 0.19\arcsec$. 
We adopt bins of 1.7 kpc (0\farcs2 arcseconds) in deprojected radius, corresponding to the MSA slit width. 
We stack the spectra in each bin and fit the emission profiles to calculate the radially averaged flux ratios across the galaxy. 
We verified that the fit residuals are in good agreement with the estimated error spectra.
Figure \ref{fig:BPT_diagram} shows the resulting BPT diagram measured for both radial bins and individual bright spaxels throughout the galaxy. For individual spaxels shown in the BPT diagram in Figure~\ref{fig:BPT_diagram} we adopt a threshold of S/N$\geq$5 in all requisite lines, exclude spaxels near the edges and those far outside the boundary of the galaxy as indicated by F444W photometric contours, and exclude spaxels with velocity or dispersion measurements near the fitting boundaries (indicating spurious fits). 
% We overlaid our data with the Sloan Digital Sky Survey (SDSS) galaxies from \cite{kewley2006}. The star-forming (SF) galaxy $z\sim0$ locus can be approximated by the functional form $O3 = a/(N2+ b) + c$ where the values for a, b and c were taken from fits by \cite{kewley2013} with $91\%$ of the distribution varying between $\pm0.1\,$dex.
%Masking pixels for emission lines are determined by the following criteria: Cutoff at SNR > 5, NAN pixels & pixels near the edges of the 2D maps (typically have a constant,nonphysical SNR value), Pixels far away from the photmetric contours defining the continuum boundary of galaxy (NIRCAM), Pixels that have velocity and dispersion measurements "near" the upper/lower fitting boundaries (i.e., difference of  <10eps)

Our data show that nebular emission within this galaxy is dominated by star formation, as all spatially-resolved regions lie well within the star-forming locus of the BPT diagram. The galaxy follows the star-forming locus within radial bins, and we can also clearly verify that prominent ``clumps'' of \Ha\ emission seen in Figure~\ref{fig:int_spec} are indeed driven by active star formation. Notably the PSF of JWST allows us to resolve the central $R\lesssim 0.8$~kpc with minimal beam smearing from emission at larger radii. We see no evidence of AGN contribution even within this central resolution element. We can place a rough limit on the contribution of nuclear emission driven by supermassive black hole accretion: assuming a central AGN with a flux ratio log(\Oiii/\Hb)~$=0.5$, a contribution of 20\% of the Balmer line fluxes from such an AGN in the central resolution element would be readily detected (approximately tripling the \Oiii\ flux relative to the star-forming locus in the central resolution element). 
The central bin contains only %$\simeq$2\% 
$\simeq$3.8\% of the total observed \Ha\ flux (although a larger fraction of total intrinsic flux accounting for attenuation; see Section~\ref{sec:attenuation}). Taking 20\% of this, we arrive at a conservative estimate that $\lesssim 0.8$\% of the total observed \Ha\ flux textcolor{magenta}{could} be associated with an AGN. The corresponding limit on the central spatial resolution element is approximately half of this ($\lesssim 0.4$\%). We conclude that either there is very little supermassive black hole accretion in this galaxy, or that any accretion-driven luminosity must be heavily dust-obscured.
This sensitive limit highlights the power of high angular resolution observations to detect low-luminosity AGN \citep[e.g.,][]{wright2010} even in galaxies dominated by star formation. 

% We do also see a collection of measurements that deviate towards the AGN branch, indicating a possible subdominant contribution of AGN \textbf{[or, more plausibly, shocks??]} to the \Ha\ flux.
% \textcolor{red}{[comment further, or remove?]}

%\textcolor{red}{[add a figure showing 1-D spectra from radial bins?]}

\subsection{Gas-Phase Metallicity Gradient}
\label{sec:metallicity}

We now turn to the gas-phase metallicity as a function of radius using the same bins described in Section~\ref{sec:BPT_diagram}. We estimate metallicity using the locally calibrated strong-line relations of \citet[][PP04]{pettinipagel2004} and \citet[][C17]{curti2017} using the available \Nii/\Ha\ and \Oiii/\Hb\ line ratios. Specifically we consider the 
$$\mathrm{N2} = \log(\mathrm{[N~II]~\lambda6584/H\alpha}),$$ 
$$\mathrm{R3} = \log(\mathrm{[O~III]~\lambda5007/H\beta}),~\mathrm{and}$$ 
$$\mathrm{O3N2} = \log \left(\frac{\mathrm{[O~III]~\lambda5007/H\beta}}{\mathrm{[N~II]~\lambda6584/H\alpha}} \right)
= \mathrm{R3 - N2}$$ 
diagnostics. 
While numerous other calibrations are available in the literature, including some based directly on galaxies at similar $z\sim1$ \citep{jones2015_metallicity,sanders2020}, the adopted calibration has little effect on the conclusions of this section. We use PP04 and C17 in order to illustrate both the numerical differences and qualitatively similar conclusions. 

\begin{figure}
    \centering
    \includegraphics[width=\linewidth]{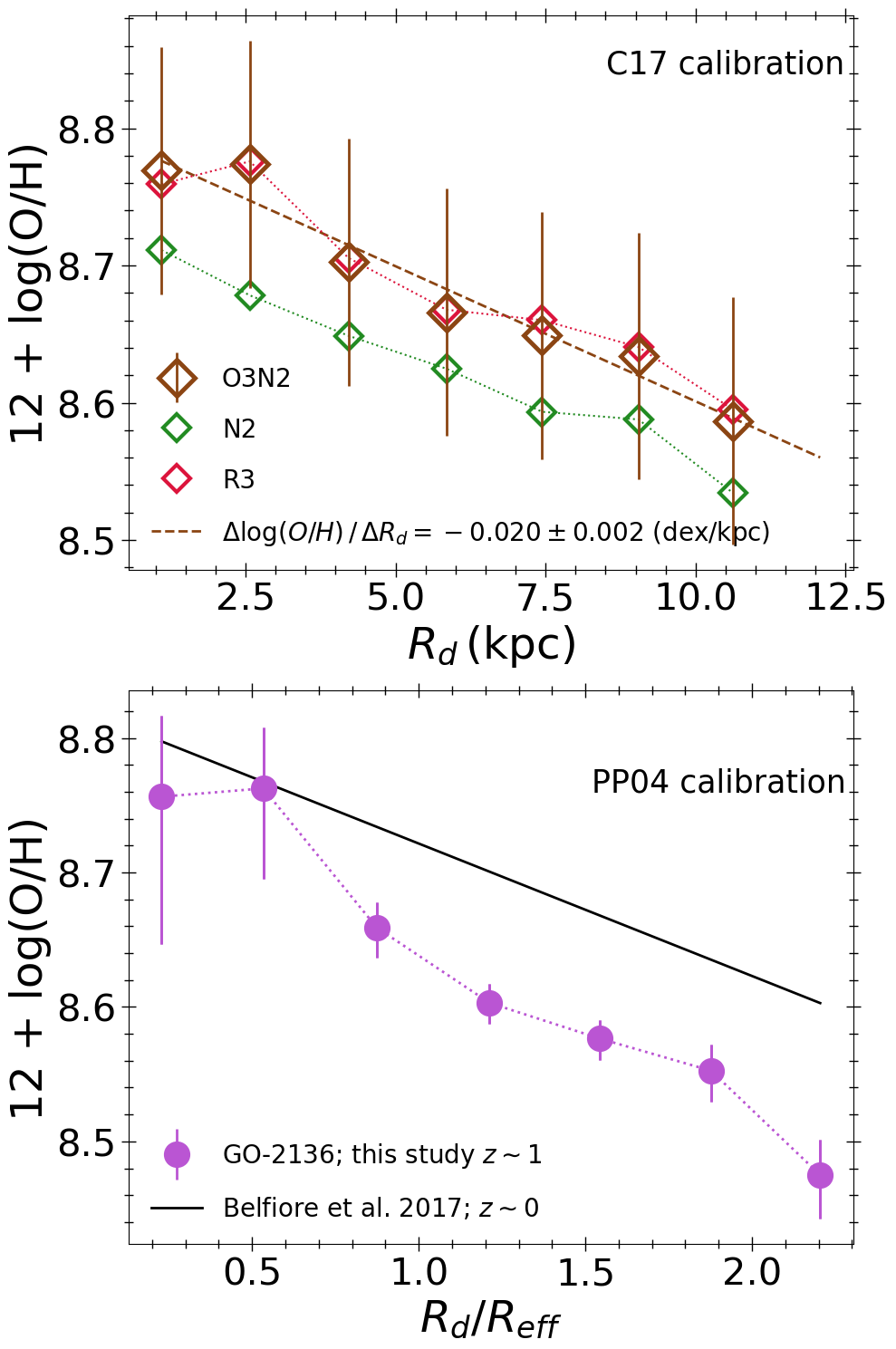}
    \caption{Gas-phase metallicity as a function of radius, showing a clear radial metallicity gradient in the target galaxy. 
    The \textit{upper panel} presents the metallicity gradient based on three different strong-line diagnostics from \citet[][C17]{curti2017} along with the best-fit gradient for O3N2 as described in the text. We observe similar gradients in each diagnostic, including the independent N2 and R3, with differences smaller than the scatter in the calibrations. 
    As an example, error bars show the 0.09 dex scatter in the O3N2 calibration.
    The \textit{lower panel} shows the metallicity gradient using the O3N2 calibration from \citet[][PP04]{pettinipagel2004}, with radius normalized to $R_{eff}$ as opposed to physical radius in the upper panel. 
    Error bars show the 1$\sigma$ statistical uncertainty (neglecting scatter in the calibration), and we note that relative uncertainty in the other diagnostics is smaller than for O3N2.
    We compare this result directly to the average gradients of $z\sim0$ disk galaxies of similar stellar mass measured by \citet{belfiore2017}. Relative to nearby galaxies at fixed mass, the $z\sim1$ target galaxy has a steeper gradient (in units of dex~$R_{eff}^{-1}$) and similar central metallicity. 
    }
    \label{fig:metallicity}
\end{figure}

In Figure~\ref{fig:metallicity} we show the gas-phase metallicity $12 + \log(\mathrm{O/H})$
as a function of deprojected radius (using the mean radius of spaxels within each bin). The top panel plots results from all three diagnostics (O3N2, N2, and R3) using the C17 calibrations. All show similar negative gradients, reflecting the radial trends in emission line flux ratios apparent in Figures~\ref{fig:radial_flux} and \ref{fig:BPT_diagram}. We additionally show a linear fit for the O3N2 index, which gives a best-fit gradient slope $\Delta \log\mathrm{(O/H)} / \Delta R_d = -0.020 \pm 0.002$~dex\,kpc$^{-1}$. 
The independent N2-based measurement differs systematically in normalization by $\sim0.05$ dex, while R3-based measurement is consistent with the O3N2-based measurement. Both are within the known scatter in these relations \citep[e.g.,][]{kewley2019}, and display a consistent gradient slope. 
%The independent N2- and R3-based measurements differ systematically \textcolor{magenta}{in normalization (N2 by $\sim0.05$ dex), which is within the known scatter in these relations \citep[e.g.,][]{kewley2019}, but display a consistent gradient slope}. 
Independent line ratios thus give a consistent overall metallicity and radial gradient.

In the lower panel of Figure~\ref{fig:metallicity} we make a direct comparison with typical metallicity gradients measured for $z\sim0$ spiral galaxies by \citet{belfiore2017}, at the \textit{same stellar mass range} ($\log{\mathrm{M_*}/\Msun} = 10$--10.5) as our target galaxy ($\log{\mathrm{M_*}/\Msun} = 10.3$). 
For consistency we compare results from the same metallicity calibration (i.e., O3N2 from PP04), normalized to the effective radius $R_d/R_{eff}$ (with $R_{eff} = 4.82$~kpc for our target; Section~\ref{sec:SBprofile}).
The PP04 calibration gives a larger total O/H range and steeper metallicity gradient compared to C17, while we emphasize the conclusion of a negative radial gradient remains robust.

\begin{figure*}[ht]
    \begin{subfigure}{.5\textwidth}
      \centering
      \includegraphics[width=\linewidth]{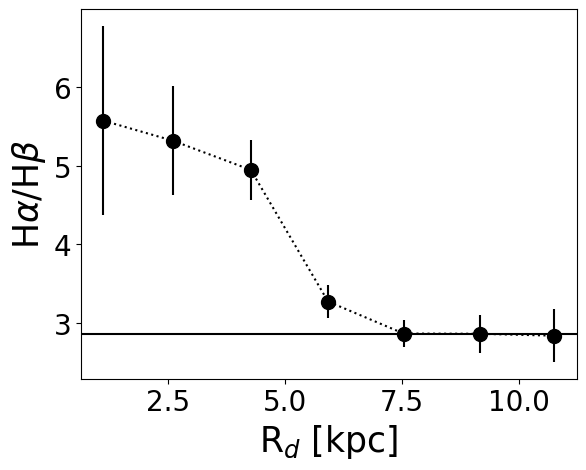}
      %\label{fig:attenuation}
    \end{subfigure}%
    \begin{subfigure}{.5\textwidth}
      \centering
      \includegraphics[width=\linewidth]{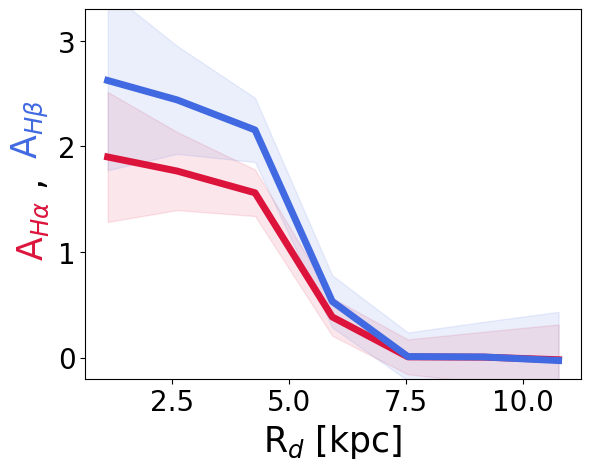}
      %\label{fig:surface_brightness}
    \end{subfigure}
    \caption{The left panel shows the measured radial Balmer decrement profile (defined as the ratio of \Ha\ and \Hb\ emission line fluxes), which is used to derive dust attenuation profiles for \Ha\ and \Hb\ (red and blue curves, respectively, in the right panel). A negative radial trend in the Balmer decrement is evident, with high values in the inner $\lesssim4$~kpc indicating a significant level of optical obscuration in these regions. In contrast the outer disk at radii $\sim$10~kpc is consistent with no attenuation.}
\label{fig:Balmer_dec}
\end{figure*}

The observed metallicity gradient is in accordance with the kinematic results which indicate a well-ordered rotating disk. Gas mixing due to turbulence, mergers, feedback-driven galactic fountains, or other factors is expected to reduce or eliminate any radial gradient \citep[e.g.,][]{ma2017,rich2012,kewley2010}. The strong gradient in this galaxy implies little influence from such effects. 
At fixed stellar mass $\log{\mathrm{M_*}/\Msun} = 10.3$, the metallicity gradient of our example galaxy at $z=1.1$ is steeper (more negative slope) than the average at $z\sim0$ from \citet{belfiore2017} while both have similar central metallicities. However we caution that the $z=1.1$ galaxy is a case study and is not necessarily representative of the general population. 
Its SFR places it below the main sequence at $z\sim1$ (Figure~\ref{fig:color_image}), such that stellar feedback and other gas-mixing process may be relatively modest.
%Its effective radius is relatively large, and 
Preliminary analysis indicates that it is in the steepest quartile of dex~$R_{eff}^{-1}$ gradient slopes within the sample observed by this program (Ju et al., \textit{in prep}). We also note that this is not an evolutionary comparison, as our target galaxy is expected to grow more massive by $z\sim0$. We leave a detailed analysis of our full sample and its chemical evolution for future work; here we have demonstrated the ability to measure metallicity gradients at a precision suitable for comparison with $z\sim0$ galaxies.

\subsection{Dust Attenuation Gradient}
\label{sec:attenuation}

Optical \Hi\ emission lines are a valuable diagnostic for characterizing dust attenuation in galaxies. The Balmer decrement, defined as the ratio of \Ha/\Hb, is a commonly used metric to determine the degree of attenuation and reddening \citep[e.g.,][]{charlot2000}. The attenuation is determined by quantifying its deviation from the theoretical value \citep{baker1938,osterbrock2006}.
%This metric is defined as the ratio of \Ha\ to \Hb\ integrated emission line fluxes, where a degree of attenuation is determined as the deviation of the observed line ratio from the theoretical value (Baker \& Menzel 1938, Osterbrock \& Ferland 2005). 
%Theoretically derived Balmer decrement value is dependent on both the electron temperature and electron density. It can vary by approx 0.3 for electron temperatures ranging between $\sim$10$^3$K and 10$^4$K, the variation with electron density is considerably less significant ($\sim$0.05; Osterbrock \& Ferland 2005). 
Assuming Case B recombination with electron temperature 10$^4$~K and electron density of 10$^{2}$~cm$^{-3}$, the expected Balmer decrement value is 2.86 \citep{osterbrock2006,storey1995}. In each radial bin, we measure the integrated \Ha/\Hb\ flux ratios from the best-fit line profiles. Figure~\ref{fig:Balmer_dec} shows the resulting Balmer decrement profile of the galaxy. It displays a clear negative radial trend, indicating larger attenuation in the central regions. %, with highest attenuation in the centroid that falls off towards the outer radial bins.

%\begin{figure}
%    \centering
%    \includegraphics[width=\linewidth]{balmer_radial_may7.png}
%    \caption{Balmer decrement radial profile}
%    \label{fig:Balmer_radial}
%\end{figure}

We infer attenuation A$_{H\alpha}$ and A$_{H\beta}$ from the measured Balmer decrement following standard methods. The attenuation is related to the color excess E(B-V) as A$_\lambda = \mathrm{k}_\lambda$ $\times$ E(B-V), where k$_\lambda$ corresponds to the reddening curve. The color excess E(B-V) can be measured from:
\begin{equation}
\begin{split}
    E(B-V) & =  \frac{E(H\beta - H\alpha)}{(k_{H\beta} - k_{H\alpha})} \\
    & = \frac{2.5}{{(k_{H\beta} - k_{H\alpha})}} \log_{10} \frac{(H\alpha/H\beta)_{obs}}{2.86}
\end{split}    
\end{equation}
where 2.86 is the assumed intrinsic \Ha/\Hb\ flux ratio. 
It is necessary to assume a reddening law for k$_{\lambda}$. For consistency with previous studies \citep[e.g.,][]{nelson2016_dust}, we adopt the \cite{calzetti2000} reddening law (with k$_{H\alpha}$ = 3.33 and k$_{H\beta}$ = 4.60). The resulting attenuation versus radius for both \Ha\ and \Hb\ are shown in the right panel of Figure \ref{fig:Balmer_dec}. 
We note that the choice of reddening law can affect the derived attenuation (e.g., A$_{H\alpha}$) -- although this effect is minor in the optical wavelength range, whereas the color excess and radial profile shape are relatively robust. We find a strong radial trend with higher attenuation in the central $\lesssim4$~kpc. This is similar to the results for massive $z\sim2$ disk galaxies found by \cite{tacchella2018} using other methods to determine attenuation.
% It is important to note that the measured color excess and, consequently, the measured dust attenuation will depend on the assumed reddening curve. (e.g. Cardelli MW is another valid choice) \textbf{[explain how this affects the results]}.

\begin{figure}
    \centering
    \includegraphics[width=\linewidth]{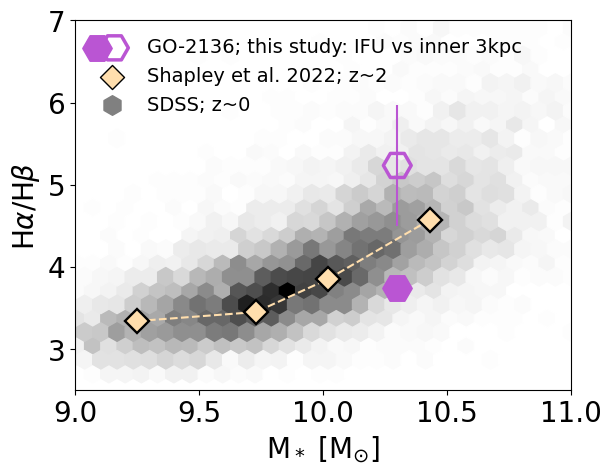}
    \caption{The Balmer decrement as a function of stellar mass at different redshifts. We measure the Balmer decrement from the integrated spectrum of our target $z=1.1$ galaxy (filled violet hexagon) and from the inner 3 kpc (open hexagon). 
    We emphasize the significance of using data from the entire galaxy to determine the Balmer decrement, as relying solely on the central region tends to yield higher attenuation estimates due to radial gradients (Figure~\ref{fig:Balmer_dec}). Our overall attenuation measurements are similar to previous findings up to $z\sim2.5$. \\
    }
    \label{fig:Balmer_mass}
\end{figure}

\begin{figure*}
    \begin{subfigure}{.5\textwidth}
      \centering
      \includegraphics[width=\linewidth]{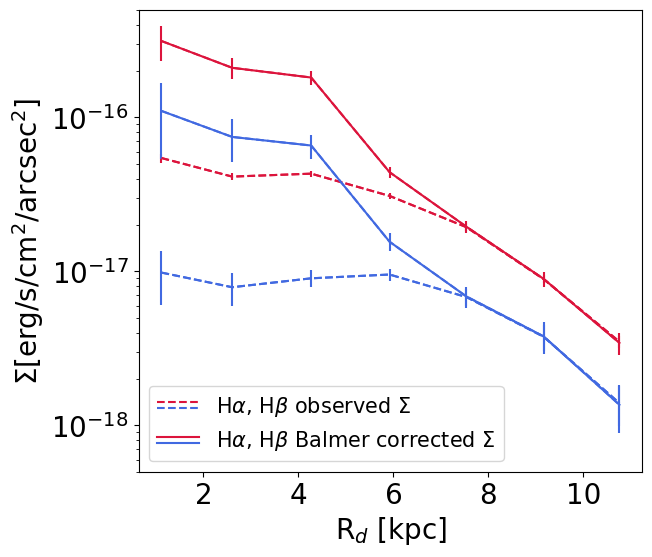}
      %\label{fig:attenuation}
    \end{subfigure}%
    \begin{subfigure}{.5\textwidth}
      \centering
      \includegraphics[width=0.93\linewidth]{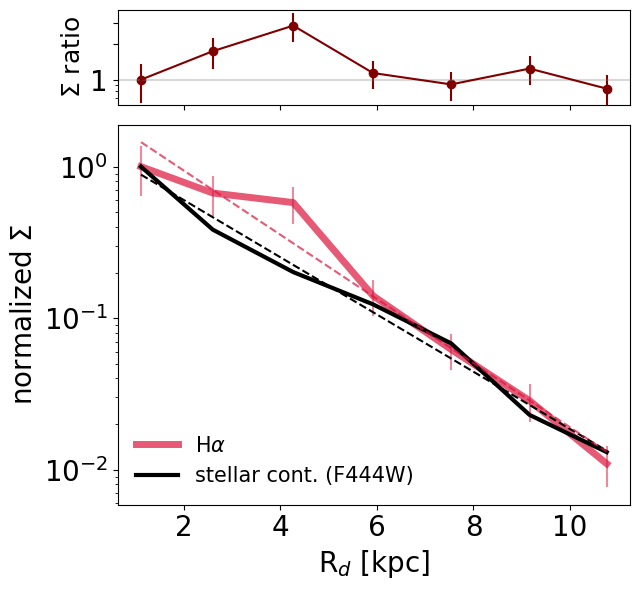}
      %\label{fig:surface_brightness}
    \end{subfigure}
\caption{Radial surface brightness profiles $\Sigma$ of \Ha\ and \Hb\ (red and blue curves respectively) are shown in the left panel for both observed (dashed) and dust-corrected emission (solid). These profiles exhibit an approximately exponential decreasing trend with radius, with an apparent excess around 4~kpc. In the right panel, we compare the \Ha\ (red solid line) and infrared stellar continuum (black solid line: NIRCam F444W filter) surface brightness profiles, normalized at their respective central bins. Exponential fits for both profiles are overlaid (dashed lines).
The upper right panel presents their ratio, which corresponds approximately to the specific star formation rate. We see a clear \Ha\ excess at $R_d \simeq 4$~kpc with nearly constant values elsewhere out to $\gtrsim 10$~kpc. 
The similar scale lengths indicate that the star formation (traced by \Ha) is not driving a rapid size growth of this galaxy. Instead it is adding mass to the disk approximately in proportion to the local stellar mass (traced by stellar continuum). 
}
\label{fig:att_sb}
\end{figure*}

Previous studies of the Balmer decrement in galaxies have noted a positive correlation with stellar mass, suggesting that more massive galaxies typically have larger attenuation from dust in their interstellar medium \citep[e.g.][]{vdokkum2005,Ly2012, kashino2013, dominguez2013, price2014, mclure2018, shapley2022}. These studies also indicate no significant evolution in dust attenuation %(as indicated by Balmer decrement) 
at fixed mass up to $z\lesssim2$. Moreover, alternative assessments of dust attenuation through indicators such as infrared excess IRX, optical attenuation A$_V$, and/or ultraviolet attenuation A$_{1600}$ and spectral slope $\beta$ have reported consistent findings \citep[e.g.,][]{wuyts2013,heinis2014,bouwens2016,bourne2017,whitaker2017,mclure2018}. 
In Figure~\ref{fig:Balmer_mass} we compare the Balmer decrement of our target from both the integrated spectrum and from the central $R_d < 3$~kpc (to mimic single-slit or fiber observations), with the results from galaxies at lower and higher redshifts ($z\simeq0$: SDSS; $z\simeq2.3$: MOSDEF, \citealt{shapley2022}).
There is a striking difference between the integrated spectrum and the central $R_d < 3$~kpc, with the central Balmer decrement being nearly 1.5$\times$ the integrated value. 
This central region is representative of the area covered by a single NIRSpec slitlet (Figure~\ref{fig:color_image}). 
This comparison shows that single-slit or otherwise centrally concentrated observations can lead to an inaccurate assessment of the overall dust attenuation in the galaxy.
 % The large difference revealed in Figure~\ref{fig:Balmer_mass} indicates that single-slit or otherwise centrally concentrated observations can potentially cause a substantial overestimate of the Balmer decrement compared to observations that encompass the entire galaxy. This example shows us that relying solely on the central region of galaxies leads to an inaccurate assessment of the overall dust attenuation in the galaxy.
The integrated measurement of our galaxy %(both whole and central slit) 
exhibits a smaller Balmer decrement compared to the average observed at a fixed stellar mass in both the SDSS sample and the $z\simeq2.3$ MOSDEF sample, although it is within the scatter seen in SDSS. 
\cite{lorenz2023} found no significant correlation between the Balmer decrement and galaxy viewing angle, suggesting that the face-on inclination of our target is not responsible for the moderately low attenuation. However they also noted a trend of attenuation increasing with star formation rate. Our target galaxy has a SFR of 3~$\Msun$~yr$^{-1}$ which places it below the typical sSFR at its redshift (e.g., \citealt{whitaker2012}; Figure~\ref{fig:color_image}), which may partly explain the lower attenuation. 
 % Lorenz+23 found no significant correlation between the Balmer decrement and axis ratio at a fixed stellar mass for their MOSDEF sample, although they noted a trend where higher masses correlate with higher star formation rates, and  higher Balmer decrements. Our target galaxy with a star formation rate of approximately (log(SFR)) 0.5 displays a Balmer decrement lower compared to the SDSS/MOSDEF running medians at a fixed mass.
%We note however that this is single data point, and a follow up study we will put our primary focus on analyzing Balmer decrement - stellar mass plane in more detail for the whole sample, and examine trends (*if any) with galaxy orientation.

\subsection{Surface Brightness Profile}
\label{sec:SBprofile}

We now examine the surface density profile of star formation, as traced by attenuation-corrected \Ha\ emission, and compare it with the stellar continuum. This provides insight into the structural evolution and radial size growth of this galaxy. 
From the measurements described in Section~\ref{sec:attenuation}, we obtain intrinsic \Ha\ and \Hb\ fluxes as a function of radius via 
$$ \mathrm{F}_{intr,\lambda} = \mathrm{F}_{obs,\lambda} \times 10^{(0.4 A_{\lambda})}. $$
Figure~\ref{fig:att_sb} shows the observed and intrinsic (i.e., attenuation corrected) radial surface brightness profiles for both \Ha\ and \Hb.
The observed profiles are relatively flat in the inner 5~kpc, with larger Balmer decrements indicating more attenuation in the central regions (demonstrated in Figure~\ref{fig:Balmer_dec}). 
The average spatial profiles found at $z\sim1.4$ by \citet{nelson2016_dust} similarly show a flattening of the inner \Hb\ profile at high masses, with larger Balmer decrements in the central regions. However the stacked profiles in \citet{nelson2016_dust} exhibit centrally peaked \Ha\ profiles across all mass bins, in contrast to the flatter inner profile seen in Figure~\ref{fig:att_sb}. 
The attenuation-corrected profiles of our target galaxy do show a clear central peak, with an approximately exponential surface brightness profile. This is indicative of star formation continuing to build an exponential disk. We find no indication of significant in situ bulge formation (in which case we would expect an elevated central surface brightness relative to the overall exponential profile), nor of inside-out quenching (in which case we would expect a lower corrected central surface brightness; e.g., \citealt{tacchella2018,lin2019}).

% Observed \Ha\ and \Hb\ surface brightness profiles indicate lower concentration towards the inner 5kpc. %The former observation is different compared to e.g. Nelson et al 2016, where they based on stacking z$\sim$1 galaxies observed \Ha\ to be more and \Hb\ to be less centrally concentrated. 
% For a slightly larger radial bin size, we do observe \Hb\ to be slightly more centrally suppressed than \Ha, consistent with the expectation that shorter wavelengths are more attenuated by dust grains. However, we still see suppression in the inner kpc for \Ha, unlike Nelson+16 where for z$\sim$1 galaxy stacks in the highest mass bin they observe \Ha\ to be more centrally concentrated, while \Hb\ is centrally suppressed. We do emphasize that their results are based on stacks in a wide mass range (9.8 $<$ $\Msun$ $<$ 11) and different inclinations, while we examine the trends for a *individual* galaxy. In the highest mass bin relevant for our case study (9.8 $<$ $\Msun$ $<$ 11) their results do suggest overall higher dust attenuation towards the center of galaxies, which is in agreement to the radial dust attenuation trend we observe for this target, with a higher dust concentration in the inner 5kpc. Dust corrected surface brightness profiles indicate both emission lines to be centrally peaked, consistent with an exponential profile, and indicative of active star-formation in the inner-most region.

Our target is covered by JWST/NIRCam imaging from the CEERS survey \citep{finkelstein2023}, which we use to trace the stellar surface brightness profile. Of particular interest is the F444W band, which measures stellar continuum at a rest-frame of 2.1 $\mu$m (i.e., rest-frame K band). We use this as a proxy for the stellar mass surface density, as the mass-to-light ratio varies by only a factor $\lesssim2$ at K band \citep[e.g.,][]{bell2001,mcgaugh2014}, which is minimal given the factor 100$\times$ in surface brightness probed here. The Balmer decrement profile (Figure~\ref{fig:Balmer_dec}) additionally suggests relatively little dust attenuation A$_K$ at these wavelengths. 
The F444W image is first rescaled to the same pixel size as the NIRSpec data cube, after which we extract the radial surface brightness profile. 
In Figure~\ref{fig:att_sb} we compare the dust corrected \Ha\ surface brightness profile (which traces the SFR) to that of rest-frame K band stellar continuum (which traces stellar mass). Both are normalized at the central bin for ease of comparison. The \Ha\ and stellar continuum both follow an approximately exponential profile with similar scale length, decreasing by a factor of $\sim 100\times$ over $\sim$10~kpc in radius. 
An exponential profile fit gives scale length $R_s = 2.9 \pm 0.3$~kpc for the continuum, corresponding to a half-light or effective radius of $R_{eff} = 4.8 \pm 0.5$~kpc.
% [note: Nelson report r$_s$(\Ha)=1.78\pm0.09~kpc for the mass bin [10-10.5]M$_{\odot}$ and r$_s$(\Ha)=2.6~kpc for [10.5-11]M$_{\odot}$; the latter would be consistent within uncertainties and is $\sim$9\% higher]
% \textcolor{orange}{The \Ha\ scale length is smaller by $11\pm9$\%, compatible within the uncertainties.} 
 % The \Ha\ and continuum are both $\sim$50\% larger than the averages reported by \cite{nelson2016} in the stellar mass bin [10 - 10.5] M$_{\odot}$.}
This is consistent with the size-mass relationship for late type galaxies at $z\sim1$, which indicate typical $R_{eff} \simeq 4$--5~kpc at the mass of our target \citep{vanderwel2014}. Additionally we note the half-light radius of our target galaxy is similar to that of the Milky Way ($\sim$4--6~kpc; \citealt{bland-hawthorn2016,lian2024}).

The \Ha\ (= SFR) and continuum (= stellar mass) scale lengths are close and mutually consistent within 1.2$\sigma$, with \Ha\ being $11 \pm 9$\%\ smaller than the continuum. 
% Ha based scale length is 2.05\pm0.18, while the continuum based -- for the same cutout as the cube, is 2.29\pm0.11
At face value this is in contrast to the picture of ``inside-out'' disk growth whereby the \Ha\ scale length is larger than that of the stars. For example, \cite{nelson2016} report that \Ha\ scale lengths of their sample of $z\sim1$ galaxies are on average 14\% larger than the observed 1.4~\um\ continuum for similar redshift and stellar mass ($\log M_*/\Msun = 10$--10.5), although $R_{eff}$ is only 4\% larger and consistent within uncertainties. While our measurements are compatible with this level inside-out growth at the $\sim 2\sigma$ level, overall it appears that the ongoing star formation is not significantly altering the galaxy's scale length. We conclude that this galaxy stellar component is growing proportionally more massive on average at all radii.

Figure \ref{fig:att_sb} also shows the ratio of \Ha\ and stellar continuum radial surface brightness profiles, normalized at the central bin. 
This ratio is proportional to the specific SFR (sSFR). 
The most notable deviation from a constant sSFR profile is a clear excess of \Ha\ emission around 4~kpc, while elsewhere the profiles align closely. The excess reaches up to three times higher \Ha\ surface brightness relative to the continuum.
% The upper panel shows the \Ha\ to stellar continuum ratio, highlighting the \Ha\ emission excess in inner radii of up to three times compared to stellar continuum.
%Figure \ref{fig:att_sb}b) illustrates radial surface brightness profiles derived from \Ha\ (red curve) and the stellar continuum from NIRCam F444W (black curve), both normalized by their respective centroid surface brightness values. We observe a pronounced excess emission in the \Ha\ profile around a radius of 4 kpc. Beyond 5kpc, surface brightness profiles of both \Ha\ and stellar continuum are mutually consistent. For clarity, upper panel of the Figure \ref{fig:att_sb}b) shows the ratio of \Ha\ to stellar continuum surface brightness profile, highlighting that \Ha\ emission in inner radii is up to three times higher than expected from stellar continuum.

One possible explanation for this \Ha\ excess emission may be prominent clumps within the spiral arms, including a bright clump in the upper left and a fainter clump in the upper right quadrant in Figure~\ref{fig:int_spec}. 
However, the Balmer decrement indicates significant dust obscuration in the inner 4~kpc, such that the morphology seen in Figure~\ref{fig:int_spec} differs from the more centrally-concentrated profile we obtain after attenuation correction. 
%We initially hypothesize that the origin of \Ha\ surface brightness excess at 4 kpc originates from prominent \Ha\ clumps within the apparent spiral arm feature — as shown in Figure 2. Specifically, we note a particularly bright clump of \Ha\ emission in the upper left quadrant of the cube cutout map in Figure 2, and a fainter clump in the upper right quadrant. However, Balmer decrement — used as the dust attenuation tracer, indicates significant dust obscuration in the inner radii. This suggests that the bright \Ha\ clumps visible in the cube cutout map likely correspond to regions with lower levels of dust obscuration along the line of sight. In contrast, \Ha\ emission in remaining central regions, particularly within the inner 4 kpc, appears to be largely obscured by dust. %Even with significant level of dust obscuration in inner radii, the star formation could still be clumpy in this apparent spiral structure. 
To further explore the nature of the excess emission at 4~kpc, we split the cube into ``upper'' and ``lower'' halves. After correcting for dust, the \Ha\ surface brightness density in the lower half is higher than that of the upper half (which contains the brightest observed regions). We find excess in attenuation-corrected \Ha\ surface brightness profiles in both halves, indicating that it is not caused solely by the discrete bright regions. Instead it may suggest the presence of a Lindblad resonance driving a ring of star formation at this radius \citep[e.g.,][]{buta1996}.

%To investigate the dominant source of the surface brightness excess at 4 kpc, we divide the cube in an “upper” and “lower” quadrant. We measure the Balmer decrement across radial bins in each quadrant to asses the level of dust obscuration. Subsequently, we derive dust attenuation corrected radial \Ha\ surface brightness profiles. Our findings reveal an \Ha\ excess at 4 kpc in both quadrants — consistent with the excess observed for the whole cube. This suggests that the origin of this excess cannot be solely attributed to the brightest \Ha\ clumps in the upper quadrant, potentially indicating the presence of a resonance ring around this radius.
 % Notably, the Balmer decrement in the innermost 4 kpc of the lower quadrant is more than twice that of the upper quadrant. Due to significant dust obscuration, \Ha\ and \Hb\ flux measurements are faint increasing uncertainty in Balmer decrement. After correcting for dust, the \Ha\ surface brightness density in the lower quadrant is approximately double that of the upper quadrant, indicating higher \Ha\ density.}
%We emphasize that due to implied high levels of dust obscuration, measurements of \Ha\ and particularly, \Hb\ flux values are quite faint, which will result in increased uncertainty in Balmer decrement measurements. Correcting for dust obscuration, the \Ha\ surface brightness density in the lower quadrant appears roughly double that of the upper quadrant at 4 kpc, suggesting a higher \Ha\ density in the lower quadrant compared to the upper quadrant.

\section{Discussion and Conclusions} \label{sec:discussion}

We have presented the design and initial results of the MSA-3D program which obtained spatially resolved spectroscopy for 43 star-forming galaxies at $z \sim 1$, using a slit-stepping observing strategy with JWST/NIRSpec's MSA. 
We illustrate the data quality with a case study of a spiral disk galaxy at $z=1.104$. This galaxy exhibits features typical of late-type spirals in the nearby universe: prominent spiral arm morphology (Figures~\ref{fig:color_image}, \ref{fig:int_spec}), rotation-dominated kinematics (Section~\ref{sec:kinematics}; Figure~\ref{fig:Ha_maps}), a negative gas-phase metallicity gradient (Section~\ref{sec:metallicity}; Figure~\ref{fig:metallicity}), and exponential surface brightness profiles in both stellar continuum and star formation (Section~\ref{sec:SBprofile}; Figure~\ref{fig:radial_flux}). 
These results collectively demonstrate the ability of our slit-stepping methodology to characterize $z\gtrsim1$ galaxies via emission line mapping with $\sim$1~kpc resolution. 
We emphasize that the galaxy analyzed in detail in this paper is typical of the data quality within the larger sample. We refer readers to Ju et al. (\textit{in prep}) for analysis of the metallicity gradients measured for 26 of our 43 targets, as well as maps of the \Ha\ flux and kinematics. 

\begin{figure*}[ht]
    \centering
    \includegraphics[width=\textwidth]{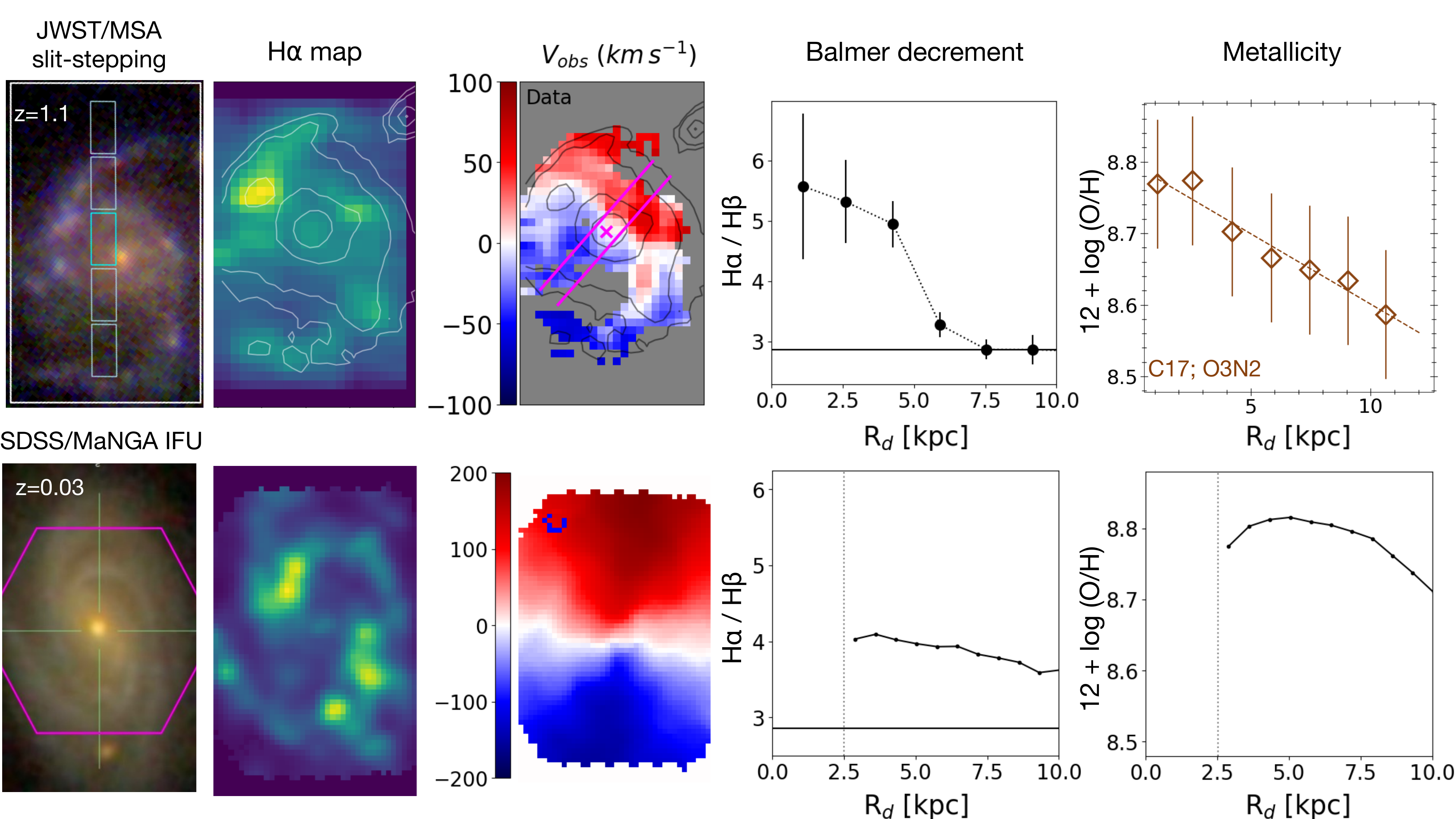} 
    \caption{
    Summary of physical measurements for the $z=1.104$ galaxy presented in this work (\textit{top}), compared to similar measurements for a galaxy at $z=0.029$ obtained from the SDSS MaNGA survey (\textit{bottom}). The galaxies are matched in stellar mass with both having $M_* = 10^{10.3}~\Msun$. Both exhibit prominent spiral structure although the $z\sim0$ galaxy is more highly inclined on the sky. 
    From left to right, each row shows: the color rest-frame optical image, \Ha\ flux map, \Ha-based velocity field, dust attenuation gradient from the Balmer decrement, and metallicity gradient measured from the O3N2 index. 
    The $z\sim0$ galaxy contains a clear AGN based on BPT diagnostics which contaminates the inner 2--3~kpc; the central regions are not shown in the attenuation and metallicity gradient plots but may influence values at $R_d \lesssim 3$~kpc. The $z\sim1$ galaxy in contrast has no sign of nuclear supermassive black hole activity. 
    This comparison demonstrates that slit-stepping with the JWST/NIRSpec MSA is capable of efficiently delivering emission line data quality for $z>1$ galaxies which is comparable to that available for large samples from MaNGA and other surveys at $z\sim0$. The MSA slit-stepping technique thus offers a promising avenue for comprehensive 3-D spectroscopic studies of galaxy formation across cosmic time.}
    \label{fig:manga_comparison}
\end{figure*}

The quality of emission line measurements we achieve with MSA-3D at $z\sim1$ is comparable to that of MaNGA \citep{bundy2015,drory2015} and other surveys of $z \lesssim 0.1$ galaxies. In Figure~\ref{fig:manga_comparison} we make a direct comparison of our MSA-3D case study target with an example galaxy from MaNGA (MaNGA ID 12-129618, plate-IFU 7495-12704) at $z=0.029$. This MaNGA target is chosen for comparison on the basis of similar spiral morphology and stellar mass ($M_* = 10^{10.3}~\Msun$). 
Figure~\ref{fig:manga_comparison} shows the color image, \Ha\ emission map, \Ha-based gas velocity field, Balmer decrement (i.e., \Ha/\Hb\ flux ratio), and radial metallicity gradient for both the $z\sim1$ and $z\sim0$ galaxies. We note that the Balmer decrement and metallicity gradient are measured from deprojected radial bins using the same methods as described in Section~\ref{sec:results} (while the \Ha\ flux and velocity maps are taken directly from the MaNGA DAP output; \citealt{westfall2019}).
A key difference which is not shown is that the MaNGA target contains an AGN, with LINER emission identified via the BPT diagram at $R \lesssim 3$~kpc. Our target $z \sim 1$ galaxy has no detected AGN emission despite sensitive limits (e.g., $\lesssim 0.8$\% of observed \Ha\ flux; Figures~\ref{fig:radial_flux}, \ref{fig:BPT_diagram}) as described in Section~\ref{sec:BPT_diagram}. An AGN contribution at the level seen in the MaNGA target would be readily detected in our MSA-3D data. 
Other than nuclear activity, both galaxies show similar structure: spiral morphology, rotating disk kinematics, higher dust attenuation toward the central regions, and gas metallicity decreasing with radius. The data quality is able to reveal subtle quantitative differences in properties such as the radial profiles of the Balmer decrement and metallicity. 
This comparison demonstrates the ability to make meaningful and detailed comparisons of our $z \sim 1$ sample with the $z\sim0$ galaxy population. JWST/NIRSpec slit-stepping data are thus powerful for galaxy evolution studies. 

We emphasize that the MSA-3D slit-stepping data enable excellent spatially resolved measurements for \textit{individual galaxies}, such as the one studied here. This represents a significant advance over previous methods, such as the %, for example, over the 
stacking technique utilized with Hubble Space Telescope grisms, which combined high-resolution optical imaging with low-resolution GRISM spectroscopy to measure radial Balmer decrement and star formation profiles \citep{nelson2016,nelson2016_dust}. Previous state-of-the-art IFS studies at $z>1$ relied on $\gtrsim10$ hour integrations with ground-based telescopes and adaptive optics \citep[e.g.,][]{tacchella2018, genzel2020}. These deep AO studies typically focus on \Ha\ emission (and the nearby \Nii\ and \Sii), whereas our MSA-3D data cover additional key lines such as \Oiii\ and \Hb, or \Siii\ and other features depending on the redshift and slit mask placement (Table~\ref{tab:sample}). The wavelength coverage of our MSA-3D observations is largely inaccessible to current AO systems, yet it is invaluable for accurately establishing the spatially resolved dust attenuation, AGN contributions, metallicity, and other properties. 
Indeed, to our knowledge the results shown in this paper represent the most robust resolved Balmer decrement, dust-corrected \Ha\ (i.e., SFR), and metallicity maps ever obtained for a $z \gtrsim 1$ galaxy.

An important conclusion from initial analyses of our MSA-3D program is that the slit-stepping strategy with NIRSpec's MSA is a \textit{highly effective and efficient method} to observe large samples of galaxies across cosmic time. 

As discussed in Section~\ref{sec:IFU_comparison}, slit-stepping is the most cost-effective approach for IFS surveys of populations with projected number density $\gtrsim 1$~arcmin$^{-2}$, which can take advantage of NIRSpec/MSA's multiplexing capability and sensitivity. 
Slit-stepping is complementary to NIRSpec's IFU mode, which provides finer spatial sampling and is powerful for studying objects which are rare on the sky. This includes gravitational lenses \citep[e.g.,][]{rigby2023} and extremely luminous systems \citep[e.g.,][]{marshall2023}.
 % We contend that slit-stepping is the \textit{only} sensible way to carry out comprehensive IFS surveys of distant galaxies with JWST's unmatched capabilities. We acknowledge that NIRSpec's IFU mode is powerful for studying objects which are rare on the sky, such as gravitational lenses \citep[e.g.,][]{rigby2023} and extremely luminous systems \citep[e.g.,][]{marshall2023}. However, MSA slit-stepping is far more efficient than the IFU for surveying sources with projected number density $\gtrsim 1$~arcmin$^{-2}$ thanks to its sensitivity and multiplexing advantages. 
Four programs have adopted slit-stepping strategies to date. 
Our $\sim$30 hour JWST Cycle 1 program has already delivered standalone data for 43 galaxies at $z\sim1$ with exquisite sensitivity, wavelength coverage, and spatial and spectral resolution. A followup Cycle 2 program (JWST-GO-3426) by our team recently obtained comparable slit-stepping IFS data for 42 galaxies at $z\simeq3$ within a single pointing. Two programs from another team have obtained slit-stepping observations of 56 galaxies at $z=1$--5 (JWST-GO-2123), and $\sim$250 galaxies at $z\simeq3$ (JWST-GO-4291), using a different and complementary dithering strategy. 
Collectively these four programs represent an investment of just over 200 hours of JWST time allocation -- less than the GA-NIFS program which is observing $\sim$50 galaxies with the IFU mode. 

Given the remarkable efficiency of MSA slit-stepping and its demonstrated success via our MSA-3D program, we advocate for future programs to comprehensively survey the galaxy population across cosmic time using this method. The four programs carried out to date are an excellent start, yet represent a modest subset of cosmic times and galaxy properties (e.g., mass and star formation rate). 
An IFS sample size equivalent to the KMOS$^{\mathrm{3D}}$ survey \citep[739 galaxies at $z=0.6$--2.7;][]{wisnioski2019} would require approximately 600 hours at the depth of our MSA-3D program, which delivers broader wavelength coverage and $\sim$5--10 times better angular resolution than KMOS$^{\mathrm{3D}}$. Such a sample can be obtained in $\lesssim 200$ hours at the shallower depth adopted by the JWST-GO-4291 program. A sample of this size is out of reach for JWST's IFU modes in its limited mission lifetime, yet it is achievable with MSA slit-stepping. 
Despite the clear advantages, however, slit-stepping is not an officially supported observing mode and the standard data reduction pipelines are not equipped to handle it. We are mitigating this by publicly releasing our custom pipeline (Section~\ref{sec:pipeline}) and example data products. While we are already addressing the main challenges associated with data processing, we also advocate for further investment in software infrastructure to optimize the scientific return of slit-stepping surveys.

\section*{ACKNOWLEDGEMENTS}

We thank B. Rauscher for developing the NSClean algorithm and making it available. 
We thank Keerthi G. C. Vasan for their valuable input.
This work is based on observations made with the NASA/ESA/CSA James Webb Space Telescope. The data were obtained from the Mikulski Archive for Space Telescopes at the Space Telescope Science Institute, which is operated by the Association of Universities for Research in Astronomy, Inc., under NASA contract NAS 5-03127 for JWST. These observations are associated with program JWST-GO-2136. We acknowledge financial support from NASA through grant JWST-GO-2136. 
This work made use of observations and catalogs from the 3D-HST Treasury Program (GO 12177 and 12328) with the NASA/ESA Hubble Space Telescope, which is operated by the Association of Universities for Research in Astronomy, Inc., under NASA contract NAS5-26555.
XW and MJ are supported by the National Natural Science Foundation of China (grant 12373009), the CAS Project for Young Scientists in Basic Research Grant No. YSBR-062, the Fundamental Research Funds for the Central Universities, the Xiaomi Young Talents Program, and the science research grant from the China Manned Space Project. 
CAFG was supported by NSF through grants AST-2108230 and AST-2307327; by NASA through grant 21-ATP21-0036; and by STScI through grant JWST-AR-03252.001-A. J.M.E.S. acknowledges financial support from the European Research Council (ERC) Advanced Grant under the European Union’s Horizon Europe research and innovation programme (grant agreement AdG GALPHYS, No. 101055023).

\vspace{5mm}
\facilities{JWST (NIRSpec MSA)}

\bibliography{JWST_Paper}
% \begin{thebibliography}{}
% \end{thebibliography}



% End of file `sample62.tex'.

\end{document}